\documentclass[aps,preprint,prd,,nofootinbib,supersciptaddress,floatfix]{revtex4-1}
\usepackage{hyperref}
\usepackage{color}
\usepackage{graphicx}
\usepackage{amstext,amsmath,amssymb,amsfonts,bbm, amsthm}
\usepackage{fancyhdr}

\def\be{\begin{equation}}
\def\ee{\end{equation}}
\def\bes{\begin{eqnarray}}
\def\ees{\end{eqnarray}}
\def\ba{\begin{align}}
\def\ena{\end{align}}

\def\bra{\langle}
\def\ket{\rangle}

\newcommand{\p}{\partial}
\newcommand{\beq}{\begin{equation}}
\newcommand{\eeq}{\end{equation}}
\newcommand{\beqr}{\begin{eqnarray}}
\newcommand{\eeqr}{\end{eqnarray}}
\def\ba{\begin{align}}
\def\ena{\end{align}}

\begin{document}

\title{ Collapse of the wavefunction, the information paradox and  backreaction}
\date{\today}
\author{Sujoy K. Modak}
\email{smodak@ucol.mx}
\affiliation{Facultad de Ciencias -- CUICBAS, Universidad de Colima, CP 28045, Colima, Mexico}
\affiliation{KEK Theory Center, High Energy Accelerator Research Organization (KEK)\\ Tsukuba, Ibaraki 305-0801, Japan}
\author{Daniel Sudarsky}
\email{sudarsky@nucleares.unam.mx}
\affiliation{{Instituto de Ciencias Nucleares, Universidad Nacional Aut\'{o}noma de M\'{e}xico\\
Apartado Postal 70-543\\
Distrito Federal, 04510, M\'{e}xico}}

\begin{abstract}
We consider  the  black  hole information problem   within the context of collapse theories  in a scheme that allows  the incorporation of the 
 backreaction  to  the Hawking radiation. We explore the issue   in a   setting of   the two dimensional version  of black hole evaporation  known as  the  Russo-Susskind-Thorlacius model. We summarize the general ideas   based on the semiclassical  version  of Einstein's  equations and then discuss  specific  modifications that  are required  in the context of collapse theories   when applied to this model. 
\end{abstract}

\maketitle

\tableofcontents

\section{Introduction}

The  black hole information question has  been  with us   for  more that    four decades,  ever since  Hawking's  discovery that  black holes   emit thermal  radiation and therefore   evaporate,   leading  either to their complete  disappearance or to  a  small   Planck mass  scale  remnant \cite{Hawking:1974sw}.    The basic  issue   can be  best  illustrated  by considering an initial  setting where   an essentially flat space-time   in  which   a  single  quantum field is in a   pure quantum state of relative high excitation   corresponding to  a  spatial  concentration of  energy,   that when left on  its own  will  collapse  gravitationally leading to the formation of  a  black hole.  As the black hole   evaporates,  the   energy that  was initially  localized   in a  small  spatial region,   ends  up  in the form of Hawking radiation that,   for  a  much of this     evolution  must  be    almost exactly thermal \cite{Hawking:1976ra}.   The point of course is that if  this  process   ends  with the complete  evaporation of the black hole (or even if a small  remnant is left)  the overwhelming majority of the  initial energy content would   correspond to a state of the quantum field  possessing almost no information (except that  encoded  in the   radiation's  temperature)  and  it  is  very  difficult to reconcile  this  with the general expectation that in  any quantum process   the initial and  final states  should  be related by a unitary transformation, and  thus  all information encoded in the initial state  must  be  somehow  present in the final one.  The issue   of course  is far more  subtle  and 
the   above should  be taken  as  only a  approximate account of the  problem. 

 There  have  been  many  attempts to deal   with   this  conundrum,  with none of  them resulting  in  a truly satisfactory  resolution of the problem \cite{mathur09, Chakraborty:2017pmn}.   In fact there is  even a  debate  as to   the extent  to which  this is   indeed  a problem  or as  some people  like to call it a  ``paradox"   \cite{Maudlin, Okon3}.
 
 In previous works   \cite{Okon0, Okon1, Okon2} we   helped   to clarified  the  basis of the  dispute, and  proposed  a  scheme  where the  resolution of the  issue  is tied to a proposal to address  another lingering problem of theoretical physics:  the so called  measurement problem \cite{bassi03} in quantum theory.

 The  first  task   was  dealt    with \cite{Okon0, Okon1,  Okon2}  by  noting that   the true problem  arises  only when one takes the  point of view  that  a   satisfactory   theory of quantum gravity must resolve the singularity, and that,  as   a  result of  such resolution, there   will be no  need to  introduce  a new  boundary  of space-time   in the region  where the  classical   black   hole  singularity stood.  Otherwise the  problem can be   fully understood   by  noting that the  region in the  black hole  exterior,   at  late times   corresponding to those  where  most of the  energy  takes the form of thermal  Hawking  radiation,  contains  no Cauchy  hypersurfaces   and thus    any  attempt to  provide  a full description of  the quantum  state  in terms of the  quantum field  modes   in the black hole  exterior is    simply  wrongheaded.   In order to   provide   a complete  description of such late  quantum   state  one  needs to  include the modes that register on  the part of the Cauchy  hypersurface that  goes  deep into the black hole  interior, in particular one  that    treads   close to the singularity, as described  in detail in \cite{Maudlin}.
 
 The second task   was carried  out  by considering the  application of  one    particular  dynamical collapse theory  designed to address the  measurement problem in quantum theory,  to  a  simple  two  dimensional   black hole  model   known as  the Callan-Giddings-Harvey-Strominger (CGHS) model \cite{CGHS92}. The  proposal  was  then to   associate  to the intrinsic  breakdown of unitary  evolution, which is typical of these dynamical collapse theories, \cite{Pearle:76, Pearle:79,
GRW:85, GRW:86, Pearle:89, GRW:90, moreCSL} (which   were developed   to deal  with the  measurement problem in standard quantum mechanics)    all the  information loss that takes place  during the  formation and  subsequent  Hawking  evaporation of the   black hole.  The first  concrete   treatments  along this line are 
 \cite {Modak:2014qja, Modak:2014vya}.

  In those  works  we  noted that  the  treatment  at that point    left  various  issues   to be  worked  out,   and    that substantial progress in  those  would be required before the  proposal  could  be  considered to be  fully satisfactory.  Among these  issues that two most pressing ones  are the  replacement of the treatment presented,  by  one that  is fully  consistent with   relativistic  covariance,  and  to show  how   the important question of  back reaction due to  Hawking  radiation on the spacetime and vice-versa can be incorporated in such a scheme (i.e., in presence of the collapse of wavefunction type setting).  A first   step in this  direction was  accomplished   in \cite{Bedingham}   where  the   simple two dimensional  problem  is   considered   using   a relativistic  version of  collapse theories.
  
  The objective of the present  work is to  continue the research path initiated in \cite {Modak:2014qja}-\cite{Modak:2016uwr} and explore an example  where the remaining issue  of backreaction in the setting of collapse theories. For this  we  will  again consider  a  two dimensional black hole model known due to Russo-Susskind-Thorlacius (RST) \cite{rst, rst2} which presents a solution of the semiclassical (Einstein) equation in 2D.

  The paper   is organized  as follows: We start by reviewing the semi-classical CGHS model in Section \ref{cgh} and then move to the RST model in Section \ref{rstm} and discuss the  quantization of matter fields  on RST in Section \ref{quant}.  It is  important to emphasize  that   all those  sections contain  nothing  novel and  represent  just a review, which    is however needed   in order to  make sense of  what follows.  Section \ref{collapse} contains necessary ingredients for the adaptation of collapse of the wave-function in a general setting as well as for the specific case of 2D RST model.  In section \ref{hawking} we discuss the Hawking radiation and the information paradox. There are two appendices (A and B) discussing important issues related with the renormalization of the energy-momentum tensor and a specific  example  of  the treatment   of back reaction  of the space-time  metric  (and dilaton field)  to   a discrete collapse of the wave-function.


\section{Semiclassical CGHS model with backreaction}
\label{cgh}

A natural  way to incorporate backreaction effects of a quantum field on the background geometry is to modify the Einstein equations where the expectation value of the stress tensor is included on the right hand side of the  equations of motion  (E.O.M), so that, 
\beq \label{ese}
G_{ab} =  T_{ab}^{Class} + \bra{\Psi}| T_{ab} |{\Psi} \ket,
\eeq
where $G_{ab}$ is the Einstein tensor of the classical metric,  $T_{ab}^{Class}$  represents the   energy-momentum tensor of   whatever matter is  being    described  at the classical level,  and $ \bra{\Psi}| T_{ab} |{\Psi} \ket $  is the renormalized expectation value of the energy-momentum tensor  of the   matter fields that are  treated  quantum mechanically,  evaluated in  the corresponding   quantum state $|\Psi \rangle$  of such fields.

 In the  two dimensional CGHS model  with  a single freely propagating massless scalar field,  characterized  by  the action \cite{CGHS92}:
\beq
S_{CGHS} = \frac{1}{2\pi}\int{d^2x \sqrt{-g}[e^{-2\phi} (R + 4(\nabla\phi)^2 + 4\Lambda^2) - (\nabla f)^2]}, \label{ac-cghs}
\eeq
 where $\Lambda$ is a constant. The  dilaton field $\phi$  is    usually treated  classically,   and the  scalar  field  $f$  is treated  quantum mechanically.

Working in  the  conformal gauge  with  null coordinates the metric  is described  by:
\beq
ds^2 = -e^{2\rho} dx^+ dx^-. \label{cg}
\eeq
The semiclassical E.O.M  involve  now the  energy-momentum   contribution   from the classical    dilaton field  and the  cosmological constant as  well as  the  part coming from the expectation value of  the quantum field $f$. Those  take   now   the following form (with respect to the appropriate variation mentioned on the left)
\beqr
\rho &:& \hspace{2mm} e^{-2\phi} (2\partial_{x^+}\partial_{x^-}\phi - 4\partial_{x^+}\phi \partial_{x^-}\phi - \Lambda^2e^{2\rho}) - \bra{\Psi}| T_{x^+ x^-} |{\Psi} \ket  = 0 \label{gpmmb}, \\
g^{\pm\pm} &:& \hspace{2mm} e^{-2\phi} (-2\partial_{x^\pm}^2\phi + 4\partial_{x^\pm}\rho\partial_{x^\pm}\phi) +  \bra{\Psi}| T_{x^\pm x^\pm} |{\Psi} \ket = 0 \label{gppmb}, \\
\phi &:& \hspace{3mm} 2e^{-2\phi}\p_{x^+}\p_{x^-}(\rho-\phi) + \p_{x^+}\p_{x^-}e^{-2\phi} + \Lambda^2e^{2(\rho-\phi)} = 0 \label{phimb}.
\eeqr
Note that  even though  the  unperturbed  metric   has  $g^{\pm\pm}  =0$   the   general variations  do not  share this property in these   coordinates,    and their consideration results in    equation (\ref{gppmb}).
 
   In order   to solve the above differential equations, it is necessary to calculate the  expectation value of various  components  of the renormalized energy-momentum tensor in  a particular state of the quantum field denoted  by   $|{\Psi}\rangle$.
   The  state   is   usually taken   to be  the  ``in  vacuum state". We review this calculation, from  a slightly  different perspective that the  usual one, in  Appendix A.

One  interesting feature of  the equations \eqref{gpmmb}-\eqref{phimb}, is that one can write down a formal action, given by
\beq
S = S_{CGHS} + S_P,
\label{acsc}
\eeq
where $S_P$ is the Polyakov {\it effective} action \cite{vassi03}
\beq
S_{P} = -\frac{\hbar}{96\pi} \int d^2x \sqrt{-g} R \frac{1}{\Box} R,\label{pac}
\eeq
and whose variation leads to the same set of equations \eqref{gpmmb}-\eqref{phimb}. This is because, in the effective action formalism,   the expectation value of the renormalized energy-momentum tensor corresponding to the  quantum field  $\hat f$, is given by the derivative of the Polyakov term 
\beqr
-\frac{2}{\sqrt{-g}} \frac{\delta S_P}{\delta g^{ab}} &=& \langle \psi|T_{ab}|\psi\rangle \nonumber\\
&=&-\frac{\hbar}{48\pi}[\nabla_a\xi\nabla_b\xi - 2\nabla_a\nabla_b\xi + g_{ab}(2R - \frac{1}{2}\nabla_c\xi\nabla^c\xi)],
\label{et}
\eeqr
where $\xi$ is an auxiliary scalar field constrained to obey the equation $\Box\xi=R$ and $|\psi\rangle$ is the  state of the quantum (scalar) field.
We  note that   the   freedom in  the choice of the  quantum state, correspond,  in the effective  action  treatment, to the freedom of  choice of boundary conditions for the
solution   $\xi$.   We refer the interested reader to  \cite{book-ns} for more discussions on the effective action formalism.

This  is a very delicate  issue  that can  generate   serious confusion in our    approach, and  care  must be taken to ensure   one goes  back and forth  from the  two  formalism in a consistent manner. We  will have to do so  in particular if we  want to    consider the   changes    in the quantum states of the $\hat  f $ field (for which the   treatment without   the effective action is more convenient) and  at the same time consider  explicitly solving  for the  spacetime  metric  and dilaton field (for  which the  reliance on  the    effective  action  is  most suitable). We  will   explore this  issue in   detail in  section  V.A. and  appendix B. In  the meanwhile  we return to the review of the original  RST model.  
 
It has been found difficult to solve the set of differential equations \eqref{gpmmb}-\eqref{phimb} without a numerical handle. The advantage of using ``effective action formalism'' is that it allows one to play with the E.O.M without going into a rigorous quantum field theory calculation, and indeed that approach was subsequently  exploited in Russo-Susskind-Thorlacius (RST) \cite{rst}, where a local term was added in \eqref{acsc}, allowing one to solve the new semiclassical equations analytically. We review this model in the next section.  

\section{Review of the RST Model}
\label{rstm}

In the RST model a local term is added to the CGHS and Polyakov actions such that the complete action,  with a   scalar   field $f$,   which are  however, treated  via  an  effective   term ,  is given by \cite{rst, book-ns}
\beq
S = S_{CGHS} + S_{P} + S_{RST}, \label{totac}
\eeq
where $S_{CGHS}$ is given by \eqref{ac-cghs}, $S_P$ is \eqref{pac} and the local term is
\beq
S_{RST} = -\frac{\hbar}{48\pi} \int{d^2x\sqrt{-g}~\phi R},
\eeq
which adds a direct  coupling between the dilaton and the Ricci scalar. Again, the  above  scheme    should  be seen as  effectively characterizing  a model  where the  Polyakov  term  replaces quantum effects of the   massless  scalar field.  

\subsection{Equations of motion}

 Next we  present the  equations of motion  that    result  from the    model's action   \eqref{totac}.  

\noindent Varying \eqref{totac} with respect to $g^{ab}$ we obtain
\beqr
&& e^{-2\phi} [- 2\nabla_a \nabla_b \phi + \frac{1}{2} g_{ab} (- 4 (\nabla\phi)^2 + 4\nabla^2\phi + 4\Lambda^2)] \\
&=& \frac{ \hbar}{48\pi}(\nabla_a\xi \nabla_b\xi - \frac{1}{2} g_{ab} (\nabla\xi)^2) \nonumber \\
&& -\frac{N \hbar}{24\pi} (\nabla_a\nabla_b\xi - g_{ab}\Box\xi) -\frac{ \hbar}{24\pi} (\nabla_a\nabla_b\phi - g_{ab}\Box\phi),\nonumber\\
\eeqr

\noindent On the other hand the E. O. M. for $\phi$ is:
\beq
e^{-2\phi} [-2R - 8\Lambda^2 + 8 (\nabla\phi)^2 - 8\nabla^2\phi] -\frac{\hbar}{24\pi}R = 0.
\eeq

\noindent In 2D conformal gauge \label{cg} {\footnote{One needs to replace $\xi$ by solving $\Box\xi =R$, $\Box = -4e^{-2\rho} \partial_{x^+}\partial_{x^-}$ and $R = 8e^{-2\rho} \partial_{x^+}\partial_{x^-}\rho$.}} 
  the  above  equations take the  following   form (with respect to the appropriate variations indicated below):
\beqr
\rho &:& \hspace{2mm} e^{-2\phi} (2\partial_{x^+}\partial_{x^-}\phi - 4\partial_{x^+}\phi \partial_{x^-}\phi - \Lambda^2e^{2\rho}) + \nonumber\\
&&  +\frac{\hbar}{12\pi}\p_{x^+}\p_{x^-}\rho + \frac{\hbar}{24\pi} \p_{x^+}\p_{x^-}\phi = 0, \label{gpmrst}\\
g^{\pm\pm} &:&  \hspace{2mm} (e^{-2\phi}- \frac{\hbar}{48\pi}) (-2\partial_{x^{\pm}}^2\phi + 4\partial_{x^{\pm}}\rho\partial_{x^{\pm}}\phi) +  \bra{\Psi}| T_{x^{\pm\pm}} |{\Psi} \ket = 0 \label{gpprst},\\
\phi &:& \hspace{3mm} 2e^{-2\phi}\p_{x^+}\p_{x^-}(\rho-\phi) + \p_{x^+}\p_{x^-}e^{-2\phi} + \Lambda^2e^{2(\rho-\phi)} + \frac{\hbar}{24\pi}\p_{x^+}\p_{x^-}\rho = 0, \nonumber\\ \label{phirst}
\eeqr
where  the expectation values of the energy-momentum tensor are those   found in \eqref{remppf} and \eqref{rempmf}. The only choice yet to implement is the selection of a particular state $|\Psi \rangle$ to solve the above set of equations. An important feature of these equations is that if one uses \eqref{gpmrst} and \eqref{phirst} one still finds the ``free field equation'':
\beq
\partial_{x^+}\partial_{x^-}(\rho-\phi)=0, \label{frp}
\eeq
which is typical  of the  CGHS model without backreaction.  This  feature   is  what in this model   facilitates  the  finding of a specific solution  for the spacetime  geometry in presence of backreaction.

\subsection{Solving semiclassical equations}

It is convenient to 
 introduce the new variables \cite{rst}
\begin{equation}
\Omega \equiv\frac{\sqrt{\kappa}}{2}\phi+\frac{e^{-2\phi}}{\sqrt{\kappa}}, \label{trans1}
\end{equation}
\begin{equation}
\chi\equiv \sqrt{\kappa}\rho-\frac{\sqrt{\kappa}}{2}\phi+\frac{e^{-2\phi}}{\sqrt{\kappa}}, \label{trans2}
\end{equation}
where $\kappa=\frac{\hbar}{12\pi}$.

In these variables \eqref{gpmrst}-\eqref{phirst} take the following form
\beqr
\partial_{x^+}\partial_{x^-}\Omega=-\frac{\Lambda^2}{\sqrt{\kappa}}e^{\frac{2}{\sqrt{\kappa}}(\chi-\Omega)}, \label{eq1} \\
\partial_{x^+}\partial_{x^-}\chi=-\frac{\Lambda^2}{\sqrt{\kappa}}e^{\frac{2}{\sqrt{\kappa}}(\chi-\Omega)}, \label{eq2} \\
-\partial_{x^\pm}\chi\partial_{x^\pm}\chi+\sqrt{\kappa}\partial^{2}_{x^\pm}\chi+\partial_{x^\pm}\Omega\partial_{x^\pm}\Omega - \frac{\kappa}{4x^{\pm^2}}+\langle\Psi|:T_{x^\pm x^\pm}:_{in}|\Psi\rangle=0, \label{eqeff}
\eeqr
whereas the free field equation \eqref{frp} becomes
\begin{equation}
\partial_{x^+}\partial_{x^-}(\chi-\Omega)=0.
\end{equation}
The above equation allows us to write
\begin{equation}
\chi-\Omega=\frac{\sqrt{\kappa}}{2}(W_{+}(x^+)+W_{-}(x^-)), \label{wrst}
\end{equation}
where $W_{+}$ and $W_{-}$ are arbitrary functions of $x^{+}$ and $x^{-}$ respectively. Then (\ref{eq1}) and (\ref{eq2}) become
\begin{equation}
\partial_{x^+}\partial_{x^-}\chi=-\frac{\Lambda^2}{\sqrt{\kappa}}e^{W_{+}+W_{-}}
\end{equation}
and
\begin{equation}
\partial_{x^+}\partial_{x^-}\Omega=-\frac{\Lambda^2}{\sqrt{\kappa}}e^{W_{+}+W_{-}}.
\end{equation}
In the RST model one restricts oneself to the choice $\Omega = \chi$, i.e., $W_{+} = 0 = W_{-}$ and then the solution is found to be
\begin{eqnarray}
\Omega= \chi = D-\frac{\Lambda^2 x^+ x^-}{\sqrt{\kappa}}-\frac{F(x^+)+G(x^-)}{\sqrt{\kappa}}, \label{soln}
\end{eqnarray}
where $D$ is an arbitrary constant and the functions $F(x^+)$, $G(x^-)$ can be found by substituting \eqref{soln} in \eqref{eqeff} and integrating

\begin{equation}\label{solnF}
F(x^+) = \int^{x^+}dx'^+\int^{x'^+}dx''^+(-\frac{1}{4x^{+2}} + \langle\Psi|:T_{x^{+} x^{+}} (x''^{+}):_{in}|\Psi\rangle),
\end{equation}

 \begin{equation}\label{solnG}
G(x^-) = \int^{x^-}dx'^-\int^{x'^-}dx''^- ( - \frac{1}{4x^{-2}} + \langle\Psi|:T_{x^{-}x^{-}} (x''^{-}) :_{in}|\Psi\rangle).
\end{equation}
Now using these expressions one can find particular solutions depending on the choice for the state of the quantum field ($|\Psi\rangle$). Specifically, we will focus on those ones which correspond to solutions representing  the formation  and evolution  of  black holes.

For future convenience let us note that in new variables the Ricci scalar, $R=8e^{-2\rho}\partial_{+}\partial_{-}\rho$ turns out to be
\begin{equation}
R=\frac{8e^{-2\rho}}{\Omega'}(\partial_{+}\partial_{-}\chi-\frac{\Omega''}{\Omega'^2}\partial_{+}\Omega\partial_{-}\Omega), \label{ricci}
\end{equation}
where
\begin{equation}
\Omega'\equiv\frac{d\Omega}{d\phi}=\frac{\sqrt{\kappa}}{2}-\frac{2}{\sqrt{\kappa}}e^{-2\phi}. \label{omp}
\end{equation}

\subsection{Dynamical case of black hole formation and evaporation}\label{dynbh}

Now we will consider the case where a sharp  pulse  of  matter forms a black hole. This pulse can be well approximated by choosing $|\Psi\rangle $ to be a coherent state build on 
top of  the  {\it in}  vacuum,   corresponding to  a    wave  packet peaked  around a 
particular classical value. In particular, we only need a left moving pulse to create a black 
hole, therefore, we can chose $|\Psi\rangle = |Pulse\rangle^{L}\otimes|0_{in}\rangle^R$ 
where $  |Pulse\rangle^{L}   = \hat {\cal O} |0_{in}\rangle^{L}$  with  $ \hat {\cal O} $  a  
suitable    creation  operator  for the  sharply peaked  wave packet. In this case the state 
dependent functions turns out to be
\begin{equation}
\langle\Psi|:T_{x^+ x^+}:_{in}|\Psi\rangle=\frac{m}{\Lambda x_{0}^+}\delta(x^+ -x_{0}^+),
\end{equation}
\begin{equation}
\langle\Psi|:T_{x^- x^-}:_{in}|\Psi\rangle = 0.
\end{equation}
This choice when  used in \eqref{soln}  leads  to  the following solution
\begin{equation}
\chi=\Omega=-\frac{\Lambda^2x^+x^-}{\sqrt{\kappa}}-\frac{\sqrt{\kappa}}{4}\ln(-\Lambda^2x^+x^-)-\frac{m}{\Lambda\sqrt{\kappa}x_{0}^+}(x^+-x_{0}^+)\theta(x^+-x_{0}^+). \label{dmetr}
\end{equation}
This  solution contains  a singularity. To see this we  refer to equations \eqref{ricci} and \eqref{omp}. The  singularity occurs   when   $\Omega' = 0$ and \eqref{omp} gives $e^{-2\phi_{s}} = \frac{\kappa}{4}$. As  we have restricted ourselves to the case $\rho=\phi$, one can use the relation \eqref{trans1} (or equivalently \eqref{trans2}), to find the value of  $\Omega_s = \frac{\sqrt{\kappa}}{4}(1-\ln\frac{\kappa}{4})$ associated with the singularity. Therefore the  location of the singularity turns out to be:
\begin{equation}
-\frac{\Lambda^2x^+}{\sqrt{\kappa}}(x^- +\frac{m}{\Lambda^3x_{0}^+})-\frac{\sqrt{\kappa}}{4}\ln(-\Lambda^2x^+x^-)+\frac{m}{\Lambda\sqrt{\kappa}}=\frac{\sqrt{\kappa}}{4}(1-\ln\frac{\kappa}{4}). \label{sing}
\end{equation}
This singularity is hidden by the apparent horizon  located at $\partial_{+}\phi=0$ which is given by
\begin{equation}
-\Lambda^2x^+(x^- +\frac{m}{\Lambda^3x_{0}^+})=\frac{\kappa}{4}. \label{ap}
\end{equation}
The apparent horizon and the singularity meet at
\begin{equation}
x_{s}^+=\frac{\kappa\Lambda x_{0}^+}{4m}(e^{\frac{4m}{\kappa\Lambda}}-1),
\end{equation}
\begin{equation}
x_{s}^-=-\frac{m}{\Lambda^3x_{0}^+}\frac{1}{(1-e^{-\frac{4m}{\kappa\Lambda}})}.
\end{equation}
The physical meaning of this point is that it could be interpreted as the end point of the black hole evaporation \cite{rst}. This is confirmed by the fact that at $x^-=x_{s}^-$ the solution \eqref{dmetr} with $x^+ > x_0^+$ takes the form
\begin{eqnarray}
\chi = \Omega = -\frac{\Lambda^2x^+}{\sqrt{\kappa}}(x_{s}^- +\frac{m}{\Lambda^3x_{0}^+})-\frac{\sqrt{\kappa}}{4}\ln(-\Lambda^2x^+(x_{s}^- +\frac{m}{\Lambda^3x_{0}^+})),
\end{eqnarray}
which is nothing but the vacuum configuration, commonly known as the linear dilaton vacuum (L. D. V.).

Thus the spacetime for $x^+ > x_{s}^+$ is given by
\begin{eqnarray}
\chi=\Omega &=&-\frac{\Lambda^2x^+}{\sqrt{\kappa}}(x^- +\frac{m}{\Lambda^3x_{0}^+})+\left[-\frac{\sqrt{\kappa}}{4}\ln(-\Lambda^2x^+x^-)+\frac{m}{\Lambda\sqrt{\kappa}}\right]\theta(x_{s}^- -x^-)\nonumber\\
& - &\frac{\sqrt{\kappa}}{4}\ln(-\Lambda^2x^+(x^- +\frac{m}{\Lambda^3x_{0}^+}))\theta(x^- -x_{s}^-). \label{dmetr2}
\end{eqnarray}

\begin{figure}[h]
\centering
\includegraphics[scale=0.5]{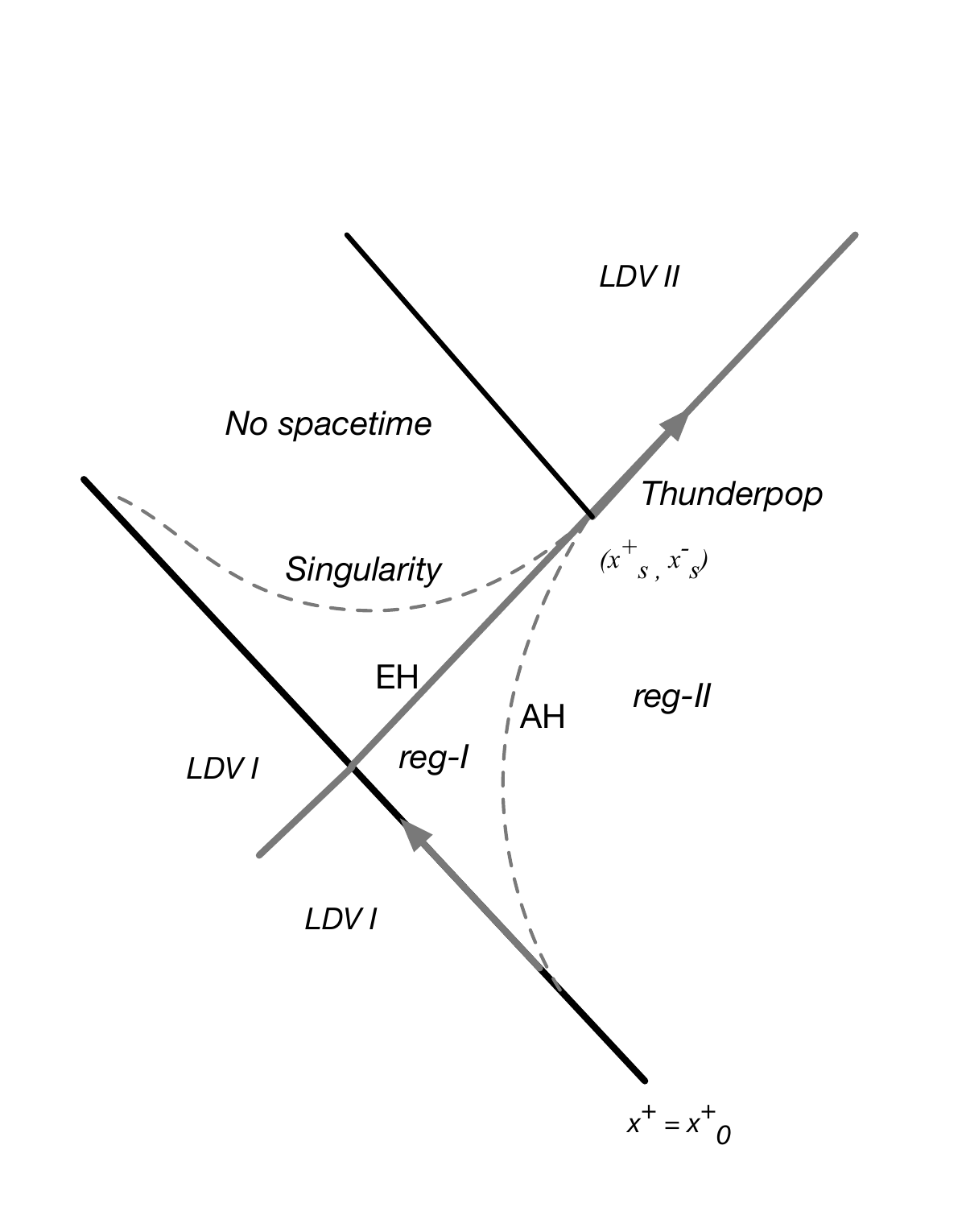}
\caption{RST spacetime in Kruskal coordinates where a black hole is created due to the matter collapse and evaporated due to the Hawking effect.}
\label{rst}
\end{figure}

Now we can construct the complete spacetime metric so that for $x^+ < x_s^+$ one has \eqref{dmetr} and for $x^+ \geq x_s^+$ the appropriate expression of the metric is given by \eqref{dmetr2}. We show the overall spacetime in Kruskal coordinates in Fig. \ref{rst}. Note that there are two different linear dilaton vacuums-- (i) for $x^+ <x_0^+$ and (ii) for $x^+ \geq x_{s}^{+},~ x^- \geq x_s^-$. These L. D. V. s are glued together with the black hole regions (Reg. I and II) by the pulse of matter (for L. D. V.-I) and radiation (for L. D. V.-II). The pulse of the matter at $x^+ =x_0^+$ carries positive energy and forms the black hole, whereas, the pulse of the radiation at $x^- = x_s^- , ~ x^+ \geq x_s^+$ carries negative energy associated with the singularity and usually called {\it thunderpop}. The space-time metric, although, is continuous at those gluing points but clearly it is not differentiable.

The Penrose diagram of the RST spacetime can be constructed following ref. \cite{Marthur}. The asymptotic past and future regions are specified with respect to the Minkowskian coordinates. First, in the asymptotic past, where the metric corresponds to a linear dilaton vacuum, so that, $ds^2 = \frac{1}{\Lambda^2 x^+ x^-}$, one can use the coordinates, 
\beqr
y^+ &=& \frac{1}{\Lambda} \ln(\Lambda x^+) - y_0^+,  \label{yp}\\
y^- &=& -\frac{1}{\Lambda} \ln(-\Lambda x^-) \label{ym}
\eeqr
to write $ds^2 = -dy^+ dy^-$, where $y_0^+$ is introduced to set the origin of the coordinates $y^+$. 

\begin{figure}[h]
\centering
\includegraphics[scale=0.5]{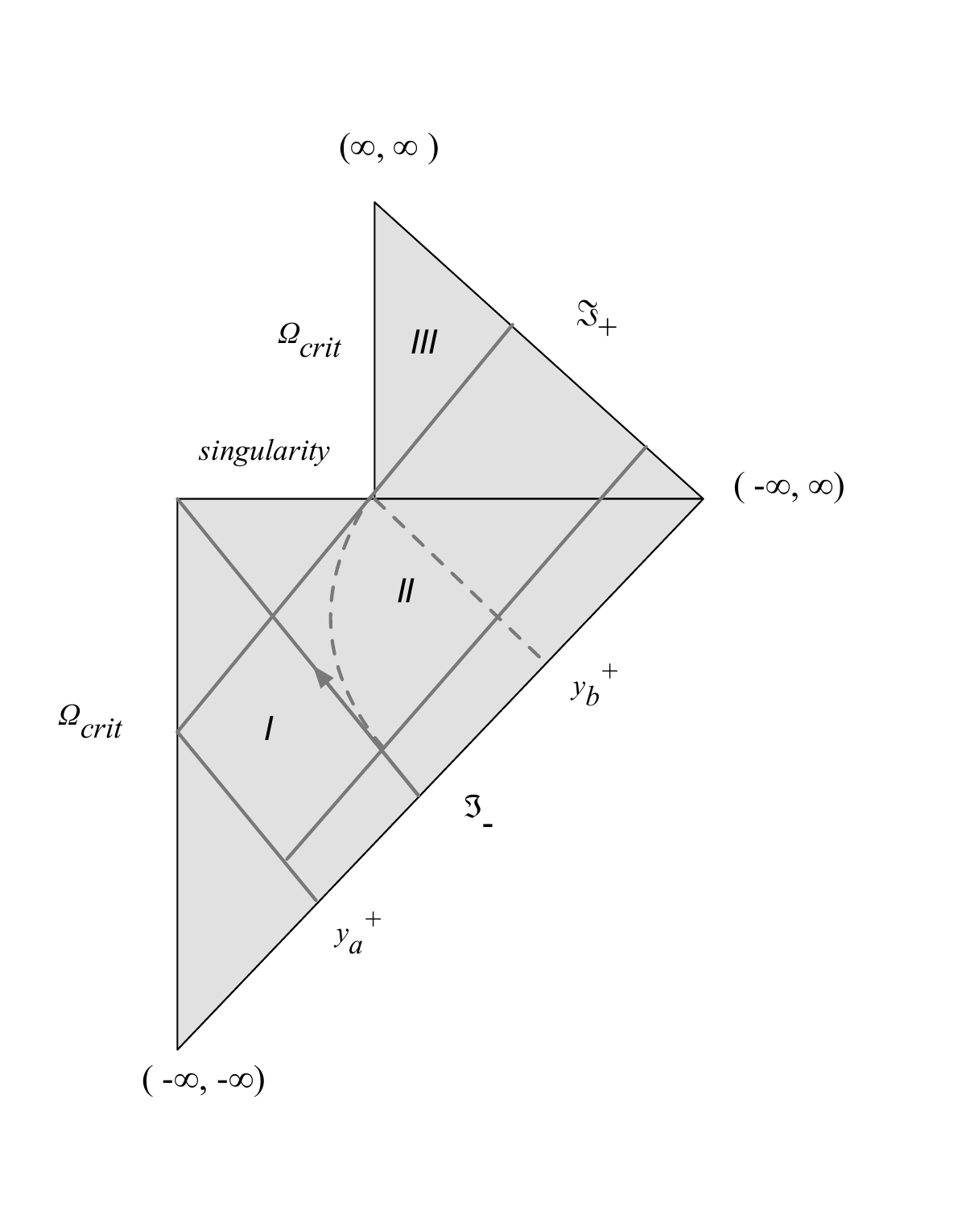}
\caption{Penrose diagram for RST spacetime where a black hole is created due to the matter collapse and evaporated due to the Hawking effect.}
\label{rst2}
\end{figure}

There is also a subtle issue regarding the extensions of the linear dilaton vacuum regions. The expression for the for the Ricci scalar (\ref{ricci}) implies that at $\Omega'=0$ it diverges. Since even in the linear dilaton vacuum regions this value can be reached one has to put some boundary conditions so that such an artifact does not show up in the solution \cite{rst}. In the literature this issue is bypassed by putting reflecting boundary conditions there. The conformal/Penrose diagram for the RST model is given by Fig. \ref{rst2}.  For a discussion about the boundary conditions to make finite curvature in $\Omega_{crit}$ see \cite{rst2}.

Let us now consider  the asymptotic structure of the spacetime. Particularly we want to check the asymptotic behavior and viability of defining ${\cal J}^+$. Let us focus on the metric in Reg. I and Reg. II (inside and outside the black hole apparent horizon), given by
\begin{eqnarray}
\chi = \Omega= -\frac{\Lambda^2 x^+}{\sqrt{\kappa}} (x^- + \frac{m}{\Lambda^3 x_0^+}) - \frac{\sqrt{\kappa}}{4}\ln(-\Lambda^2 x^+ x^-) + \frac{m}{\Lambda\sqrt{\kappa}}. \label{metr23}
\end{eqnarray}
If we want to find out the physical metric coefficient in the conformal gauge \eqref{cg} we need to use the relations \eqref{trans1} and \eqref{trans2}. By using those we obtain the following equation
\begin{equation}
\frac{\sqrt{\kappa}}{2}\rho+\frac{e^{-2\rho}}{\sqrt{\kappa}} - \Omega = 0, \label{solrho}
\end{equation}
whose solution determines $\rho$. However, practically this equation is not invertible and therefore we cannot find an exact solution for $\rho$ from the known expression of $\Omega$. But we can perform certain analysis to unfold the asymptotic behavior. First, we check the staticity of the metric by expressing \eqref{metr23} in the standard Schwarzshild like coordinates ($t,~r$). These are related with Kruskal ($x^{\pm}$) coordinates in the following way
\begin{eqnarray}
x^+ &=& \frac{1}{\Lambda} e^{\Lambda \left(t + \frac{1}{2\Lambda}\ln(e^{2\Lambda r} - \frac{m}{\Lambda})\right)}, \label{transks1} \\
x^- &=& -\frac{m}{\Lambda^3 x_0^+} - \frac{1}{\Lambda} e^{-\Lambda \left(t- \frac{1}{2\Lambda}\ln(e^{2\Lambda r} - \frac{m}{\Lambda})\right)}. \label{transks2}
\end{eqnarray}
Using these relations the metric function takes the following form
\begin{eqnarray}
\chi = \Omega &=& \frac{1}{\sqrt{\kappa}} e^{2\Lambda\sigma} - \frac{\sqrt{\kappa}}{2} \Lambda\sigma - \frac{\sqrt{k}}{4} \ln(1 + \frac{m}{\Lambda^2 x_0^+} e^{\Lambda (t - \sigma)}), \label{metrsch} \\
\sigma &=& \frac{1}{2\Lambda} \ln(e^{2\Lambda r - \frac{m}{\Lambda}}).
\end{eqnarray}
Now note that for $t= const.$ and $r \rightarrow \infty$ (i.e., at spatial infinity $i_0$) the last term which is time dependent vanishes altogether. This makes $\Omega$ time independent and therefore any solution for $\rho$ in \eqref{solrho} will be time independent. This guaranties the staticity of the metric at $i_0$. Furthermore, as one moves up in ${\cal J}_R^+$ where $t \rightarrow \infty, ~ \sigma \rightarrow \infty$ keeping $t-\sigma = \textrm{finite}$, the time-dependent term becomes least dominant and one can approximately define an asymptotic time-like Killing vector field near ${\cal J}_R^+$. We shall use this asymptotic Killing time to be associated with the physical observers as usually done in the RST model. To talk about the asymptotic flatness let us focus on \eqref{metrsch}. Near ${\cal J}_R^+$ as $\sigma,~t,~x^+$ all tends to infinity, from \eqref{metr23} and \eqref{solrho} and, by comparing the dominating coefficients of $\kappa$, we have $\lim_{x^+\rightarrow\infty} e^{2\rho} = \frac{1}{-\Lambda^2 x^+ (x^- + \frac{m}{\Lambda^3 x_0^+})}$, which is the asymptotic form of the LDV-II. Thus in entire ${\cal J}^+_R$ one has a flat metric. This is essential for discussions related with Hawking radiation and information paradox.

Up  to this  point  we have presented the standard  RST  model  which   incorporates the   backreaction of the  spacetime to the  Hawking  evaporation, corresponding to the  usual   quantum   evolution of   the matter field,  and  which  indicates that information is  either  lost   at the singularity or somehow  conveyed  to the exterior in  some  unknown  fashion encoded  in the  very late outward  flux of  energy known as  the {\it thunderpop} .

 However,   in  the context of our proposal,  the  state of the  quantum field  will be   affected  by the    modified   quantum   evolution prescribed by  the modified dynamics involving spontaneous collapse of the wavefunction. The specific model that we shall consider is given by the Continuous-Spontaneous-Collapse (CSL) theory, a brief introduction of which will be provided in section V.  In light of this modification the backreaction of the quantum matter on the spacetime  metric  will  be modified  as well.  We  will discuss  this in the   reminder of this  work .

\section{Quantization on RST}
\label{quant}

In order to  discuss in  some  detail   the  modifications,  brought in  by  the CSL  version of quantum  theory,  it is  convenient  to   describe the  two  relevant  constructions of  the quantum theory  of the scalar field  $f$ on RST spacetime.

We  note,  however that the   power of this  model  resides in the fact that one  is  able to obtain the     whole space-time,  including the  back reaction   of the   space-time   metric to  the quantum  energy momentum  stress tensor,    before  one  actually  discusses the construction of the quantum    field theory for the matter   field  $\hat f$.  This,  in turn, allows  for that construction to be  carried out in the   appropriate    spacetime   which    already  includes backreaction.

 Thus  one  might think that  one can safely  ignore  this  part  of the treatment and just go ahead  with the  usage of the effective action  and  never actually carry out the explicit construction of the quantum field. This would be correct  except that   in our approach we   will need to  further consider  the changes in the state of the quantum field   brought about  by  the dynamical collapse theory.   Doing that   requires   the quantum   field theory   for   $\hat f$   and  we proceed  with this now.

  We can express the scalar field both in the {\it in} region and the {\it out} region. The {\it out} region consists of the modes having support in  the inside and outside of the event horizon. Since in the discussion related with Hawking radiation one only considers the modes in the ``right moving'' sector we shall only use them in various expressions here. In the {\it in} region  one can write
 \begin{equation}
f=\sum_{\omega}(a_{\omega}u_{\omega}+a_{\omega}^{\dagger}u_{\omega}^{*}),
\end{equation}
where the {\it in} vacuum is defined by $a_{\omega}|0\rangle_{in}=0$. In the {\it out} region one has
\begin{equation}
f=\sum_{\omega'}(b_{\omega'}v_{\omega'}+b_{\omega'}^{\dagger}v_{\omega'}^{*}) +\sum_{\tilde{\omega}'}(c_{\tilde{\omega}'}\tilde{v}_{\tilde{\omega}'}+c_{\tilde{\omega}'}^{\dagger}\tilde{v}_{\tilde{\omega}'}^{*}),
\end{equation}
where the modes with and without tildes respectively have supports inside and outside the horizon. Vacuum within the Fock spaces interior and exterior to black hole are respectively $c_{\tilde{\omega}'}|0\rangle_{int} = 0$ and $b_{\omega}|0\rangle_{ext} = 0$. Using Bogolyubov coefficients one can express the creation or annihilation operators of the {\it in} region in terms of a linear combination of creation and annihilation operators (defined in either Fock spaces) of the out region. Specifically, there are two sets of Bogolyubov coefficients connecting the {\it in} region to black hole interior and exterior regions (only for the {\it right moving sector} of the scalar field modes). 

The field modes appearing in above expressions of $f$ are respectively given by
\beqr
u_{\omega} &=& \frac{1}{\sqrt{2\omega}}e^{-i\omega y^{+}}, \\
v_{\omega'} &=& \frac{1}{\sqrt{2\omega'}}e^{-i\omega'\sigma^{-}},
\eeqr
where $y^{+}$ is defined in \eqref{yp} and $\sigma^{-}=-\frac{1}{\Lambda}\ln\left(\frac{\Lambda x^{-}+\pi M/\Lambda\kappa}{\Lambda x_{s}^{-}+\pi M/\Lambda\kappa}\right)$ \cite{Marthur}. Whereas the mode in the interior of the black hole can be defined from the expression of $v_{\omega}$, in the following way
\begin{equation}
\widetilde{v}_{\tilde{\omega}'}(y^-)=v_{-\omega'}^{*}(-y^-),
\end{equation}
with $y^-$ given by \eqref{ym}. One should also note that in the continuous basis (using $\omega, \omega'$) modes are in fact not orthonormalized and to talk about particle creation in a particular quantum number one has to introduce a discrete basis to make sense of the particle definition. Moreover, the  discrete  basis  allows  for a relatively simple  characterization of  localization of the modes  which   in the  continuous  basis  would require the use of   wave packets.  We  will  rely  on   the discrete basis in our work.  It is easily obtained from the continuous  counterpart   by defining    the modes,
\beq\label{transd}
v_{jn}=\frac{1}{\sqrt{\epsilon}}\int_{j\epsilon}^{(j+1)\epsilon}d\omega e^{2\pi\omega n/\epsilon}v_{\omega}^{out}
\eeq
with $n$ and $j\ge 0$ are integer numbers. These wave packets are peaked about $u_{out} = 2\pi n/\epsilon$ with width $2\pi/\epsilon$. The Bogolyubov coefficients between the $u_{\omega}$ modes and $v_{\omega'}$ (and its complex conjugate) modes turns out to  satisfy the following relationship in the late time limit \cite{Marthur}
\begin{equation}\label{bogo}
\alpha_{\omega\omega'}\approx e^{-\pi\omega/\Lambda}\beta_{\omega\omega'}.
\end{equation}
It is also possible to express the Bogolyubov coefficients in the discrete basis just by using the transformation \eqref{transd}.

A standard calculation from the above expression immediately leads us to the  Hawking radiation. Also, as it is well known, with this relation one can express the {\it in} vacuum (defined in the Fock space at ${\cal J}^-_R$) as a superposition of the particle states in the joint basis in  the out region defined inside and outside the event horizon (i.e., interior to black hole and at ${\cal J}^+_R$) \cite{Giddings, giddings2}. Later, we need  to consider  the initial state of the quantum field which will be evolved using the CSL evolution,  with the CSL term  taken   as  an  interaction hamiltonian. We   will  take this state, to be  the {\it in} vacuum for right moving sector and a pulse for the left moving sector, as discussed in subsection \ref{dynbh}. Using Bogolyubov transformation we can express
\beqr
|\Psi_{i}\rangle &=& |0_{in}\rangle^R\otimes |Pulse\rangle^{L}, \nonumber\\
&&= N\sum_{{F}}C_{F}|F\rangle^{int}\otimes|F\rangle^{ext}\otimes  |Pulse\rangle^{L}\label{inqs}, \label{instate}
\eeqr 
where the {\it in} vacuum for right moving sector is expressed as  a linear combination of $|F\rangle$ states interior and exterior to the black hole. Each of these states is characterized   by a set of  excitation numbers corresponding to various modes. Entanglement between the interior and exterior modes,    in the above state, implies that corresponding to a $|F\rangle^{ext}$ there is a single  $|F\rangle^{int}$ with the same particle excitation number but with negative energy/wave-vector. It is this initial quantum state \eqref{inqs} that had led to the formation of   the black hole. We should also state that using the  late time  characterization,  the  usual  thermal nature of the radiation can be  seen in the fact  that  $C_F = e^{-\pi E_F/\Lambda}$ , where $E_F = \sum_F{\omega_{nj}F_{nj}}$ is the total  late  time  energy  of the $|F\rangle$ state.

Now that we have characterized  the initial quantum state we move to the next section to incorporate   the CSL evolution on this state.

\section{Incorporating collapse mechanism in the RST model}
\label{collapse}

We  will be   addressing the  question of the   fate of    information 
 in the evaporation of the black hole  by  considering a modified  version of quantum theory proposed  
 to address the  ``measurement problem'' of the standard quantum theory. 
 The  specific  version  that we  will be using   is  the  CSL theory proposed in \cite{Pearle:89}. This theory is a  continuous version of  the so called Ghirardi-Rimini-Weber (GRW) theory \cite{GRW:85}, \cite{GRW:86} where the unitary evolution is accompanied by occasional discrete collapse of the wavefunction that happens for a very small amount of time.  As these theories were developed  in the context 
 of    many particle  non-relativistic   quantum mechanics  we   will need to  adapt it to the present  
 context involving  quantum field theory in curved spacetime{\footnote{Ideally one would finally need to  use a fully  relativistic version of  collapse theories such as \cite{BedinghamTh} - \cite{Kay} as was done   in \cite{Bedingham} for the non-backreacting case.}}.

\subsection{Collapse of   the  quantum state  and   Einstein's  semiclassical equations}

One  of the main difficulties that  must be  dealt with   when  considering   a   semi-classical treatment of    gravitation in the context of   modified  quantum theories involving  a  collapse of the quantum state   is  the fact that  Einstein's equations  simply will not  hold   when  the  energy-momentum tensor is  replaced  by  its quantum  expectation value  and the quantum   state  of  the matter fields  undergoes  a stochastic  collapse.   In fact  this is connected to   the   intrinsic problem of   treating gravitation in  a classical  language, and  it is  expected to   be   fully 
 solved only in the context of a  complete    theory of  quantum  gravity.  This  is     illustrated  in \cite{Page}   where it is  argued  that  semi-classical  gravity  is  either   inconsistent when  we  
 assume the  state of   quantum   matter  undergoes  some  sort of collapse,  or, it  is  simply  at odds with  experiments  when  we do not make  such an assumption.   Unfortunately we  do  not have at this point a fully  workable  quantum   gravity theory  to  explore    these  issues. Furthermore,    even  when, and if   we   eventually  get  our hands  on   such  a theory,   in which   as  expected the  classical spacetime   metric is  replaced  by  some  more fundamental  set of quantum  variables,   the  recovery  the  standard  notions   of  classical  spacetime, a  task  that seems  unavoidable  if  we  want  to be  able to   describe  such  things  a  formation and  evaporation of   black holes,  can be   expected to   be  a  rather   complex  process that moreover might  only  work  in  some  approximate  sense.
 
 These considerations  lead  us to  adopt the following approach:   We  will   consider semi-classical  gravity  as  an   approximate and  effective   description,    valid  in limited   circumstances,   of a  more fundamental   theory of  quantum gravity  including  matter  fields.  This  seems to be in fact the  position  that would be adopted in this regard  by a good  segment of our community working on these questions,   but  we  describe  our posture  explicitly  in order  to  avoid misunderstandings.   The  analogy to  keep in mind  is  the hydrodynamic  description of fluids,  which   as we  know,  works  rather  well  in a    large   class of  circumstances,  but  does not   represent the  behavior  of the  truly fundamental   degrees of  freedom  involved.  We  know,  that   at  a deeper level fluids  are   made of   molecules  that interact in  a complex  manner,  and that, there are   only certain aspects  of their collective   behavior  describable in the   hydrodynamic  language.   The   semi-classical  Einstein  equations   therefore cannot be trusted to  hold precisely at the fundamental microscopic level,   just  like the Navier-Stokes   equations, which   cannot be thought to  represent  the  true behavior of  the fluid molecules,  but  must be  taken as holding only   in  an  approximate   sense.    Moreover,  just  as the hydrodynamic   characterization of a fluid   is    known to  break  down   rather   dramatically in  certain circumstances,   such as  when a  ocean  wave   breaks  at the  beach,   we  can also  expect that  Einstein's  semiclassical equations   should become invalid under some  situations.  We    thus  must take the   situations  associated   with   the   collapse  of the  quantum state   of matter fields to be  one  of such  circumstance.

   Now   we  must  consider   a way out to  formally  implement such  ideas in order to  be  able to  further  explore  them and  their consequences, and  in  particular to apply   them to the problem   at hand. 
 
 Below,   we  discuss  the    issue,  first in the realistic  setting of a  3+1  dimensional  spacetime  and second  in the    particularly simple  situation concerning the   RST  model    in 1+1    dimensions.
 
 Here  we  will   describe an   approach   initially    proposed   in \cite{Alberto}  in the context of     inflationary cosmology and  the problem of  emergence  of the primordial inhomogeneities \cite{Inflation-Collapse}.  The staring point is  the notion of   Semi classical  Self- consistent Configurations (SSC),   defined for the case of a single  matter field (for simplicity),  as  follows:
 
 Definition: The set $\lbrace g_{ab}(x),\hat{\varphi}(x), \hat{\pi}(x), {\cal H}, \vert \xi \rangle \in {\cal H}\rbrace $ represents a SSC if and only if $\hat{\varphi}(x)$, $\hat{\pi}(x)$ and $ {\cal H}$ correspond a to quantum field theory constructed over a space-time with metric $g_{ab}(x)$
and the state $\vert\xi\rangle$ in $ {\cal H}$ is such that
\begin{equation}\label{scEE}
G_{ab}[g(x)]=8\pi G\langle\xi\vert \hat{T}_{ab}[g(x),\hat{\varphi}(x)]\vert\xi
\rangle, 
\end{equation}
  where  $\langle\xi| \hat{T}_{\mu\nu}[g(x),\hat{\varphi}(x)]|\xi\rangle$  stands for the renormalized  energy momentum  tensor   of the quantum matter field $\hat{\varphi}(x)$ (in the state $\vert\xi\rangle$) constructed   with the space-time   metric $  g_{ab}$.    This corresponds, in a sense, to the general  relativistic   version of Schr\"odinger-Newton equation \cite{Diosi:1984xi} - \cite{SN1}. The  point  of  this  setting   is  to ensure  a consistency  between the   description    of the   quantum  matter and  that  of gravitation by considering their influences on each other\footnote{We   consider this  as a scheme  to be   used   when  all matter    is  treated   quantum mechanically, but  one  might  add the   contribution    to the    energy-momentum  tensor   from   any    fields   which are treated  classically, such as  the dilaton  in the RST  model.}.  

To this setting we want to add an extra element: {\it the collapse of the wave function}.  That is, besides the unitary evolution describing the change in time of the state of a quantum field,  we 
consider    situations   such as  those  envisaged  in    discrete  collapse theories   such as  GRW, where  there will be, sometimes, spontaneous jumps in the quantum state.
We  will    consider  the situation when  we are given a    dynamical   collapse theory that, given an SSC (dubbed  as SSC1  and  considered to  describe the situation before the collapse),   specifies,   a  space-like  hypersurface  $\Sigma_{Collapse}$ (perhaps  through some   stochastic  recipe that we can overlook at this point) on which the  collapse  of the  quantum state   takes place,   and  also  the   final quantum state (generally, again in a  similar stochastic  manner). The   remaining task   is   now twofold -- (i)  to describe    the   construction of  the  new SSC, (to be  called  SSC2)  that  will be  taken to describe  the situation after the collapse  and, (ii)  to  join the  two  SSC's in a manner to  generate   something to call, in its closest sense, a ``global space-time".

In order to have  a  picture  in  our mind  we  can think of the  above  scheme as  something  akin to  an   effective  description  of   a fluid   involving a situation where ``instantaneously",  the  Navier-Stokes  equations do not hold.  Let  us think, for instance,  once more,  about   an  ocean  wave  breaking  at the  beach.  The situation    just before  the wave   breaks  should be   describable  to a  very  good  approximation   by the Navier-Stokes  equations, and   should    the situation,  well  after the wave  breaks and  the water surface becomes    rather smooth again. The  particular  regime where the  breakdown of the wave   is taking place, and  its  immediate aftermath,  will, of  course, not be the one  where  fluid  description  and  the Navier Stokes equations can be  expected to  provide  an  accurate picture. This   is because  such   regime  involves    large  amounts  of  energy and information   flowing   between the macroscopic  degrees  of freedom   that are  well characterized  in the fluid   language,  and the  underlying molecular   degrees  of freedom,  (accompanied  by  other complex  process including   such things  as  incorporation of   air molecules into  the   water, the   mechanism  by  which  ocean  water is  oxigenated).  All these   represent  aspects   would that  have  been  ``averaged  out" in passing  from the  molecular     to the   fluid  description.  If  we  now take   the  limit  in  which this   complex  non-fluid   characterization  is  essential  to  be {\it instantaneous}, then we  will be in  possession of two  regimes  that  are  susceptible  to  a  fluid  description  using Navier-Stokes  equations,   joined  through an {\it instantaneous collapse of the wavefunction}  (to  be identified  with the  space-like  hypersurface $\Sigma_{Collapse}$ ) where the said equations cannot hold.  

The  specific  proposal  for the   effective  characterization of  these situations,  that   we  will  have in  mind  is   based on the  $ 3+ 1$  decomposition of the  space-time  associated with the hypersurface  $ \Sigma_{Collapse}$  and inspired by the   application  of these  ideas  in the   specific   case   treated in \cite{Alberto}. 

The spacetime  metric   of the SSC1  defined on   
$ \Sigma_{Collapse}$ has the  induced  spatial metric  $h_{ab}^{(1)}$, the  unit normal  $ n^{a (1)}$ and the   extrinsic  curvature ${K^{ab}}^{(1)}$.    
The  fact that the  SSC1  corresponds to a  semi-classical solution of Einstein's  equations  then ensures that the  Hamiltonian  and    momentum  constrains  are satisfied  on $ \Sigma_{Collapse}$  viewed as  an hypersurface  embedded  in  the  space-time  of the SSC1.

 The  task, assuming that we  are  given  the   expectation of the energy-momentum tensor   for the SSC2  is  given,  i.e.,   assuming that the  collapse theory  allows  us to    determine  
  ${\langle\xi\vert \hat{T}_{bc}[g(x),\hat{\varphi}(x)]\vert\xi}\rangle^{(2)}$. That   is the  collapse  theory,  in our case  CSL,     gives us  the  quantum  state at   any  hypersurface (assuming that the initial  state   was  given)  on  the $\Sigma_{Collapse}$  
  and   given  that    one  would need is to construct the full  SSC2.    The  first  thing  would be    obtain  suitable initial data  for the  space-time metric  of SSC2. That  is   we  need  to find     $h_{ab}^{(2)}$  and the   extrinsic  curvature ${K^{ab}}^{(2)}$  satisfying   the Hamiltonian and Momentum  constrains,  involving the  expectation    value of the energy-momentum   tensor   corresponding   to the SSC2.
  
  We  have previously  used   the  {\it  ansatz} of taking      $h_{ab}^{(2)} = h_{ab}^{(1)}$ on    $ \Sigma_{Collapse}$ and   finding   a  suitable  expression   
$K_{ab}^{(2)} $, on 
 $ \Sigma_{Collapse}$   determined so  as to  ensure the   Hamiltonian  and  momentum  constraints   for the  SSC2  are satisfied.
 
 This  determination of the initial data  for the SSC2    metric    will be referred  as  Step 2. 
 Next   one  would have to carry out  the completion of the SSC2, namely  to  specify the   construction of  the  quantum field theory,  identify the  quantum  state   and  show  how the  full  space-time   metric   can  be   determined.

The   completion of the process  would   then   involve  finding   the state of  the new   Hilbert space  such that   its  expectation  value of  the (renormalized)  energy-momentum tensor  corresponds to the  values  given     in  Step 1  above.  One   must then  ensure   that with  the  integration of the  space-time  metric      and  the  mode functions  given  those initial data can be  done  effectively.   An  explicit    example   showing the  completion of this   process in the    inflationary  cosmological context   representing a  single   mode perturbation  with   specific  co-moving   wave vector  was  presented in  \cite{Alberto}.    There  one can see that in  general  the   tasks involved are  rather non-trivial. 
    
      The point is that this   scheme  will allow the   construction of  a    space-time  made of two   four dimensional regions  characterized  by   the SSC constructions  and  joined  along a  collapse  hypersurface  where   Einstein's  equations   do not  hold.

    
In the case  of  the  RST  model,  in 1  + 1  dimensions,   the situation is  substantially simplified    by  the fact that the space-time is  two dimensional  and thus  the Einstein  tensor vanishes  identically.  The  violation of  the equations of motion  during a collapse of the quantum state   are thus  a bit more   subtle and,  at the same time  easier  to deal  with.   

 Again, one  can make use of the fact that  the most  general   spacetime   (smooth)   metric   can  be  written  as:
 \beq
ds^2 = -e^{2\rho} dx^+ dx^-, \label{cg2}
\eeq
where    $\rho  $ is a smooth  function.   Thus   both the   space-time metric  for SSC1  and  SSC2  can be put in this  form.  The issue  is now  joining these two  space-times  along  $\Sigma_{Collapse}$.
  Thus  we can regard,   as  a complete  generalized space-time, the result obtained  by this  gluing procedure  where the  price  we    have payed in so doing is that now the   function $\rho$    will not   necessarily be  a  smooth  function. Note that something similar   happens when the linear dilaton vacuum region is glued with the black hole space-time in CGHS or RST model.    
  

 Also  the  remaining    SSC 2  construction  i.e. the   specification of  the   corresponding Hilbert space  and identification of the   quantum state needs to  be  dealt with. The  specification of   the   mode functions   that  determine the Fock space, can be taken to  be done at the  level of initial data on  $ \Sigma_{Collapse}$,  and   this    can be achieved by   making use of the fact  that    the  general solution of  the   Klein-Gordon  equation   for a massless  scalar field   $f$ on   any space-time metric of the form (\ref{cg2})  is   of the form   
 \beq\label{fmodes}
 f = f_{+} ( x_+) + f_{-} ( x_-).
\eeq
 As  we  mentioned  before,  the   idea  is  then to  take  take the modes  used in the SSC1  Fock space as providing initial data  for the SSC2  Hilbert  space construction.   However  it is  easy to see    that    using   such  procedure  on $ \Sigma_{Collapse}$    with functions  satisfying (\ref{fmodes})  both   in the regions to the past  and   future of     $ \Sigma_{Collapse}$     corresponds,   simply, to functions   satisfying   (\ref{fmodes})  in  all  our    generalized   space-time  ( i.e.  after  incorporating the  discontinuity  in the  derivative  of  $\rho$).   
  All  this  will    work  fine    as  long   as    the spatial-metric    is continuous  along    $ \Sigma_{Collapse}$   and   the space-time  metric is   continuous as  one  crosses  it. 

The  point is   that    we  can take the modes   of   the SSC2  and    the corresponding Fock space to  be  the same  as those of   the SSC1.  This    represents  a   very nice  simplification   provided  by the  two  dimensional  nature  of the situation of interest.  
   
The   last   step   would  be   the    identification of  the state in the Hilbert space of SSC 2   with the appropriate   expectation values of the renormalized  energy-momentum tensor,  but again as  the   Hilbert space of   two  SSC's   are the same  we can  take this  as  being   provided  by the   collapse theory   in  Step 1. 

  This  shows that   the program  described  at the  start of this  section,  regarding a single      instantaneous  collapse  of the  state of the  quantum matter field,   can be   easily implemented   in the  present situation. The    details of the  general application of that scheme for   situations   in higher dimensions is   an open   problem. 
  
The final issue that needs to be  considered  before    applying   collapse  theories to the problem  at hand,   has  to do  with   generalizing the  above  procedure, from  the case  of a single   instantaneous collapse   to a continuous   collapse  theory.  That is,  if  instead of considering   a    single collapse   taking place on the space-like  hypersurface     $ \Sigma_{Collapse}$  we   want to consider  a   foliation of space-time  by    space-like ``collapse"  hypersurfaces   $ \Sigma_{Collapse} (\tau) $ parametrized  by a  real  valued  time   function  $ \tau$ on the  space-time manifold,  and  a  theory   like  CSL to  be  described  in the next section,   describing the   change in the  quantum  state,   as one  ``passes   form  one hypersurface to the other".  
 
  We  can deal  with  this, first for the case of  a finite  interval   $[\tau_{start}, \tau_{end}]$ in the  time  function by  considering  a partition of the  corresponding   interval   $ \tau_0=\tau_{start}, \tau_1......\tau_i...  \tau_N= \tau_{end}$ ,   performing the  procedure describing the   individual   discrete  collapse at  each  step   $i$ in the partition,  and  eventually taking the limit $N\to \infty$. 
  
With these  considerations  in hand  we  now  turn to the  description of the specific   type of  collapse theory will be   using in this  work.

\subsection{CSL Theory}
  We  first   consider the   theory in the non-relativistic quantum mechanical  setting  in which it was  first   postulated.
  The  CSL   theory is   generically  described  in   terms of  two equations.
The first is a modified  version  of Schr\"odinger equation, whose   general solution,  in the  case of a  single non-relativistic  particle is:
  \begin{equation}\label{CSL1}
|\psi,t\rangle^{CSL}_{w}={\cal T}e^{-\int_{0}^{t}dt'\big[i\hat H+\frac{1}{4\lambda}[w(t')-2\lambda\hat A]^{2}\big]}|\psi,0\rangle,
\end{equation}
where ${\cal T}$ is the time-ordering operator. $\hat A$   is   a  smeared  position operator   for the  particle.  $w(t)$ is a random classical function of time, of white noise type, whose probability is given by the second equation, the Probability Rule:
  \begin{equation}\label{CSL2}
	PDw(t)\equiv\langle\psi,t|\psi,t\rangle\prod_{t_{i}=0}^{t}\frac{dw(t_{i})}{\sqrt{ 2\pi\lambda/dt}}.
\end{equation}
\noindent The state vector norm evolves dynamically (\textit{not} equal 1), so eq. (\ref{CSL2}) says that the state vectors with largest norm are most probable.
 It is  straightforward to   see  that the total probability is 1, that   is 
 \begin{equation}\label{CSL3}
\int PDw(t) =\langle\psi,0|\psi,0\rangle=1.
\end{equation}

The   way  we will  incorporate the CSL  modifications in   our  situation  is    by  relying  on the    formalism   of interaction picture version   of quantum evolution, where   the   free part of the evolution corresponding to the standard  quantum evolution    will be absorbed in the  construction of   the  quantum field  operators, while  the interaction  corresponding to the   CSL modifications  will  be used to  evolve the  quantum  states.  One  more thing that needs to be  modified  is related to the   fact that  the   quantum field  is  a  system  with infinite  number of  degrees of freedom (DOF), and  thus instead of a single operator   $\hat A$     and a single   stochastic function $w(t)$  we will have  an infinite set of those labeled by the index  $  \alpha$. Thus  in  our case   we  will have:

  \begin{equation}\label{CSL-QFT}
|\Psi,t\rangle^{CSL}_{ \lbrace w_{\alpha} \rbrace}={\cal T}e^{-\int_{0}^{t}dt'\big[\frac{1}{4\lambda} \sum_{\alpha}
[w_{\alpha}{} (t')-2\lambda\hat A_{\alpha}]^{2}\big]}|\Psi,0\rangle,
\end{equation}
with a  corresponding  probability rule for the joint realization of the  functions  $\lbrace w_{\alpha} (t)\rbrace $. 

It is not our aim to review various physical and technical features of CSL theory here for which we refer the reader to our previous papers \cite{Modak:2014qja, Modak:2014vya} as well as well established papers and review articles in the literature \cite{bassi03}. However, we would like to add an important point that is worthy to highlight here. Since CSL theory adds a non-linear, stochastic term to the otherwise deterministic Schrodinger equation, there is an inevitable loss of information associated with it. Given an initial quantum state we cannot predict the final state after CSL evolution with 100\% accuracy even in Minkowski space-time. Given the tiny numerical value of the collapse parameter $\lambda \sim 10^{-16}sec^{-1}$ this departure is so small that no observable effect can be found in practical situations while dealing within laboratory systems and hence making the theory phenomenologically viable. Nevertheless, it is an important insight that stochasticity that was brought in by CSL theory allows information destruction in quantum evolution. The major challenge that we overcame in our earlier proposals \cite{Modak:2014qja, Modak:2014vya} was making this tiny effect substantially  larger inside a black hole  in   a sensible  manner. We review this important feature in the next subsection.   

\subsection{Gravitationally induced collapse rate}
In order to intensify the tiny loss of information controlled by $\lambda$ we made an hypothesis that $\lambda$ is a function of local curvature of the space-time. Mathematically we expressed \cite{Modak:2014qja, Modak:2014vya}
\beq
\lambda(R) = \lambda_0\left(1 + \left(\frac{R}{\mu}\right)^\gamma\right), \label{lamr}
\eeq
where $\mu$ is an appropriate scale and $\gamma\ge 1$. In flat spacetime this reduces to standard CSL theory. The advantage of the above equation is that now we can tune the rate of collapse by changing the local curvature (Ricci scalar for 2-dimension and Weyl scalar for 4-dimension). While doing that we are in fact tuning the dominance of the non-linear, stochastic term over the linear term in \eqref{CSL1}. The non-linear term helps us to break the linear superposition of various basis vectors and stochasticity brings a high degree of indeterminism in the quantum evolution and these two effects together is the cause of information destruction while singularity of the black hole is approached. Of course with our present knowledge we cannot verify or falsify the hypothesis used in \eqref{lamr}, but in future one might be able to find out a way to do this and this will, in turn, provide a strong evidence in favor or against of our proposal. We consider this to be a clear advantage over the others existing in the literature.  

Now we need to mathematically implement the above ideas in RST model and the first step towards this is to foliate the spacetime with Cauchy slices.    

\subsection{Spacetime foliation}
\label{sub-fol}

In order to  describe the evolution of quantum states we   will need to foliate the space-time with Cauchy slices and introduce  a suitable global time parameter labeling the foliation.  We perform  an analysis similar to that used in  works \cite{Modak:2014qja, Modak:2014vya} for CGHS model,  with some needed modifications. The relevant patch of the black hole space-time is referred as Region I and Region II in Fig. 1. Region I is further divided into two regions Region I(a) and I(b) which are within the event horizon and apparent horizon respectively. A Cauchy slice has the following characteristics-- $R= const.$ curve in Regions I(a) and I(b), joined with a $t=const.$, curve in Region II. Note that this time $t$, defined with respect to an asymptotic observer, is well defined in regions II and I(b), i.e., outside the event horizon. The family of slices are determined once we specify the intersecting points between the $R=const.$ and $t=const.$ curves. For that we have to find out a curve which stays within Region I(b). This is needed to ensure that the Cauchy slices are spacelike and forward driven with respect to the asymptotic Killing time $t$. There might be many such curves and any of them should be as good as others to do the job. We choose the curve
\begin{eqnarray}
-\Lambda^2 x^+ (x^- + \frac{m}{\Lambda^3 x_0^+}) - \frac{\kappa}{4}(1+ a_0 (x^- - x_s^-)(x^+ - x_0^+)) = 0, \label{int}
\end{eqnarray}
where $a_0$ is a constant and we shall use a fixed value for this in our analysis. 

Let us now comment on the quantitative aspect of making these slices. The targeted regions are I and II as shown in Figure 1. We start by calculating $R$ in \eqref{ricci} by using \eqref{metr23} which gives
\beqr
R &=& -\frac{16}{\Lambda  x^- x^+ x_0^+ \left(\kappa e^{2 \rho }-4\right)^3} \left(e^{2 \rho } \left(\Lambda  \kappa^2 x_{0}^{+} + 4 \kappa m x^+ + 16 \Lambda ^2 x^-  x^{+2} \left(m+\Lambda ^3 x^- x_0^+ \right)\right)\right. \nonumber \\
&& \left. + \kappa^2 \Lambda ^3 e^{4 \rho } x^- x^+ x_0^+ + 16 \Lambda ^3 x^- x^+ x_0^+\right). \label{rc}
\eeqr
In principle it is straightforward to find the $R = const. = R_0$ slices in the following way. First we can solve \eqref{rc} for $e^{2\rho}$ as a function of coordinates $x^\pm$ and other parameters ($\kappa, m, \Lambda, x_0^+$) since the equation $R-R_0 = 0$ is a cubic equation in $e^{2\rho}$. Among the three solutions for $e^{2\rho}$ one has to take the real solution and put it in \eqref{trans1} to find $ \Omega_c$ which now corresponds to $R=R_0$. The next step is to equate \eqref{metr23}  with $\Omega_c$ and this in turn allows one to write $x^+ = f_{R_c}(x^-)$ where $f_{R_c}(x^-)$ is a function of $x^-$ on the collapse hypersurface with curvature $R = R_c$. The resulting analytic expression for the $R=R_c$ curve is too cumbersome to put in a paper and therefore not included here.

\begin{figure}
\centering
\includegraphics[scale=0.5]{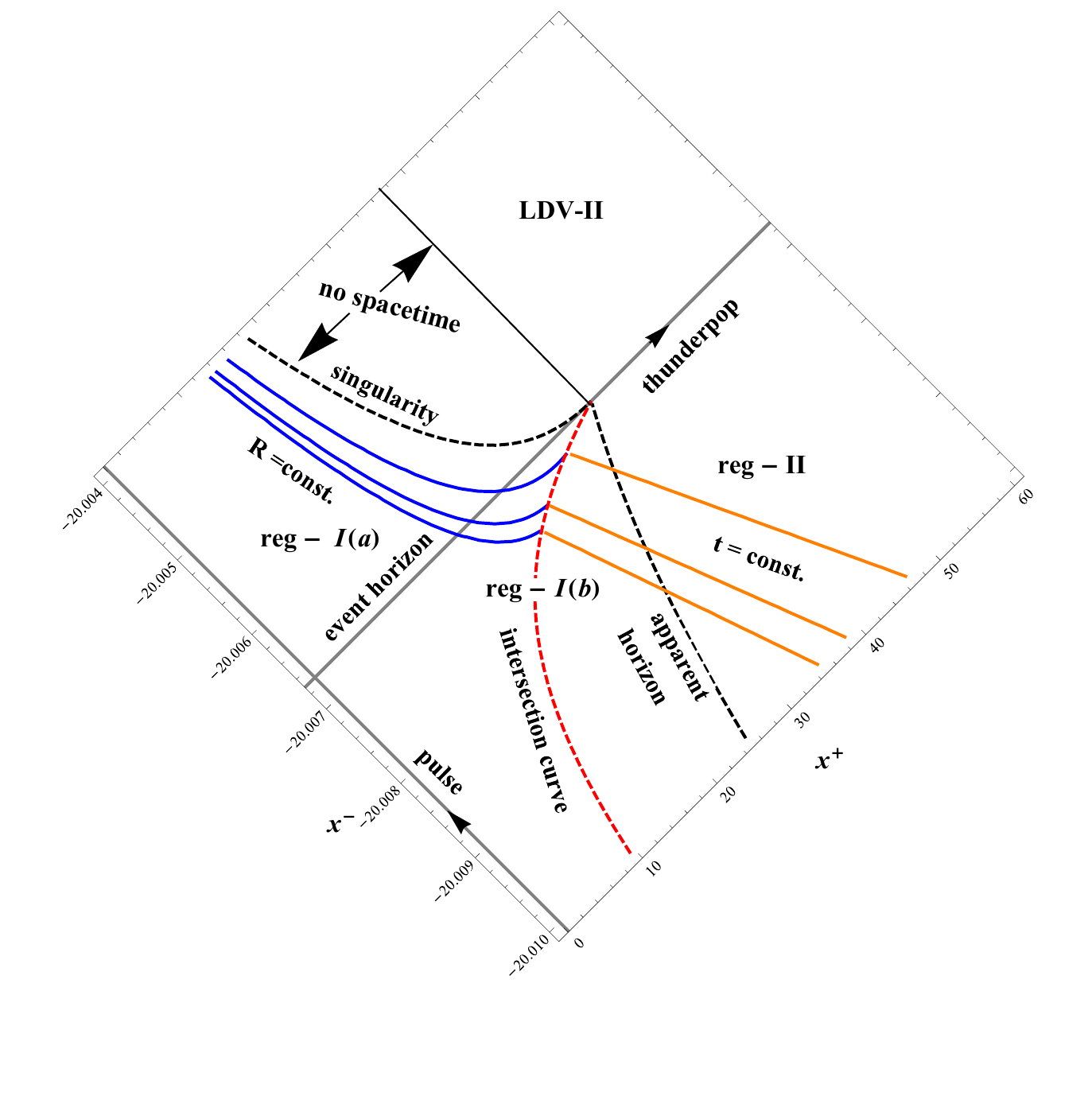}
\caption{Spacetime foliation plots for RST model. The family of Cauchy slices are given by $R=const.$ curves inside the apparent horizon joined with $t=const.$ lines outside. We have set $k=1, m=2, \Lambda =1, x_0^+ = 0.1$ for all plots. Values of curvature $R = 50, 30, 25$ for blue curves from top to bottom. As required for this slicing to be Cauchy, hypersurfaces with larger $R$ (more closer to the singularity) are joined together with larger values of asymptotic Killing time $t$.}
\label{rst-fol}
\end{figure}

However, it is possible to plot the $R= const.$ curve numerically along with other relevant curves, namely, the singularity \eqref{sing}, the apparent horizon \eqref{ap}, the intersection curve \eqref{int} and $t=const.$ lines to complete the foliation. This family of Cauchy slices is plotted in Fig. 3. It is clear that slices with increasingly higher curvature are those that are closer to the singularity. Making the collapse parameter a function of the spacetime curvature intensify the collapse of the wave function near singularity. This, in effect, erase almost all the information about the matter that had once created the black hole.

\subsection{CSL evolution and the  modified back reaction}
Now we are in a position to use the construction made so far and evolve the initial quantum state as given in \eqref{instate}. We have assumed the curvature dependent collapse rate in \eqref{lamr} and prepared our Cauchy slices in the preceding subsection. The only thing that is missing is to define the set of collapse operators appearing in \eqref{CSL-QFT}. As we have indicated before that it is convinient  to use the discrete basis in order  to have  at our  disposal   simple  notions of  localized   exitations   or ``particles''. Therefore   it is    convinient to   characterize the collapse operators  also  in  terms  of this discrete basis.  Following our earlier works,  and  for simplicity  we chose \begin{equation}\label{number}
A_{\alpha} := N_{nj}=N^{int}_{nj}\otimes\mathbb{I}^{ext}.
\end{equation}
Here $N^{int}_{n,j}$ is the number operator in the discrete basis (which was used to discretize the mode function in \eqref{transd}) interior of the black hole and $\mathbb{I}^{ext}$ is the identity in the exterior of the black hole basis. The number operator acts on the particle states as follows
\begin{equation}
N_{nj}|F\rangle^{int}\otimes|F\rangle^{ext}=F_{nj}|F\rangle^{int}\otimes|F\rangle^{ext},
\end{equation}
where $F_{nj}$ is the particle excitation in a state corresponding to quantum numbers $n,j$.

At this point we must   consider the  fact  that when we introduce the CSL  modification of the  evolution  of  the quantum   state of the field,  
we also have  a modification of  the   expectation  value  of  the renormalized energy-momentum tensor (REMT), and this  will   in turn modify the
back reaction of the quantum field on the background spacetime.

One  issue that needs to  be mentioned  here is that  such  simple  collapse operator might not  be  physically   appropriate  (or  acceptable)   as it might fail to   take    Hadamard  states into Hadamard  states.   The  issue  of what are the kind of operators   that when  used  as collapse  generating operators  would  ensure  this essential  property    has   been studied    with some  generality    in \cite{Kay}, however    the results   suggest that   there are  suitable  choices of     acceptable  operators  that    are  very close to the one  described  above.   

 In order   to consider the   modification of  the  back reaction  we  should  in principle    study the expression for the   REMT in the quantum state $|\psi_{CSL}\rangle$ which according to \eqref{rem2} can  be written in the following form: 
\beq \label{rem2d}
\langle \psi_{CSL} | T_{\mu\nu}(x)| \psi_{CSL} \rangle_R = \langle \psi_{CSL} |: T_{\mu\nu}(x) :_{in}| \psi_{CSL} \rangle + \text{Geometric terms},
\eeq
where the first term on the r.h.s is normal ordered with respect to the ``in'' quantization. Here we shall be considering the right moving sector of field modes and concentrate how the REMT evolves due to CSL evolution (for a brief account on the issue see Appendix A).

In fact  what  we    would need to  do   is to  compute  the quantities  $\langle \psi_{CSL} |: T_{\pm \pm}(x) :_{in}| \psi_{CSL} \rangle$    that  should be used  in (\ref{remppf})  or more  specifically  withing the full RST  model in (\ref{solnF}) and (\ref{solnG}).

The  {\it in}  normal  ordered    energy-momentum tensor can be  easily  expressed in terms of the  objects used  to  construct the  ``in''  quantization  region and takes the following form
\beqr\label{remin}
: T_{\pm\pm} (x) :_{in}&=& \lim_{x\rightarrow x'} : \frac{1}{2}(\partial_\pm \hat  f (x)) (\partial_\pm  \hat  f (x'))  :_{in}\nonumber\\
&=& \frac{1}{2} \int{d\omega_1} {d\omega_2} 
\lbrace \lim_{x\rightarrow x'}(\hat a_{\omega_1} \hat a_{\omega_2} u_{\omega_1, \pm} (x) u_{\omega_2, \pm}(x') +\nonumber\\
&&2\hat a_{\omega_1}^\dagger \hat a_{\omega_2} u_{\omega_1, \pm} (x)^* u_{\omega_2, \pm}(x') +
\hat a_{\omega_1}^\dagger \hat a_{\omega_2}^\dagger u_{\omega_1, \pm} (x)^* u_{\omega_2, \pm}(x')^* \rbrace, 
\eeqr
where $_{,\pm}$ represents ordinary derivative with respect to $x^{\pm}$. Note  that the  normal ordering procedure has been  explicitly performed in the above expression. Also it should be cleared that   although the modes  appearing in the above  expression  are  naturally  associated   with the   early flat     dilaton  vacuum  region    they are defined  everywhere,  so the  expression is  valid  for evaluations of the expectation  value anywhere  in  the  space-time.
 
 As we mentioned before we shall focus only on the  right moving modes for CSL effect. The changes in  the  state  due to the  CSL  modified  evolution are   expected to be relevant only  at late times,  due to the fact  that   that  is  where the parameter $\lambda$  becomes   large   due to the  large values of the curvature  as the singularity is approached. Thus in principle  we  only  need to   consider    the  CSL  modifications  of the state   after the  pulse.
 
 The state  in fact   can be  written  as 
\beqr \label{CSL-QFT3}
| \psi_{CSL} \rangle &=&{\cal T}e^{-\int_{0}^{t}dt'\big[\frac{1}{4\lambda} \sum_{nj}
[w_{nj}{} (t')-2\lambda\hat N_{nj}]^{2}\big]}
N\sum_{F}C_{F}|F\rangle^{int}\otimes|F\rangle^{ext}
\nonumber\\
&=&N \sum_{F}C_{F} e^{-\int_{0}^{t}dt'\big[\frac{1}{4\lambda} \sum_{nj}
[w_{nj}{} (t')-2\lambda F_{nj}]^{2}\big]}
|F\rangle^{int}\otimes|F\rangle^{ext}
\eeqr
with $C_F = e^{-\frac{\pi E_F}{\Lambda}}$ in the late time limit.  One of the   complications in  evaluating the  quantity of interest is that  while in the  above  expression   all the operators refer to the  {\it out}  quantization,  the   expression (\ref{remin}) uses the  {\it in}   quantization.  In principle  one  could   rewrite all the    operators  appearing in  (\ref{remin})  using the Bogolyubov   relations  and end  up  with an  expression for $\langle \psi_{CSL} |: T_{\pm \pm}(x) :_{in}| \psi_{CSL} \rangle$   where  everything is   expressed  in terms of the  {\it out}  quantization.  Furthermore  once  one  considers  a specific  realization of the   stochastic functions $\lbrace  w_{nj}\rbrace $,  one   would have a well defined  expression,   which  could, in turn,  be used to   compute the  modifications of the    spacetime  metric  given  by the functions $ F\&G$  in     (\ref{solnF}) and 
(\ref{solnG}).  That is in \eqref{remin} we can  express the creation and annihilation operators defined in the ``in'' region, by using Bogolyubov transformations, by operators associated with the   black hole {\it interior}  and the {\it exterior} region:
\beqr 
\hat a_{\omega} = \sum_{jn}\alpha_{jn,\omega} \hat b_{jn}  + \beta^*_{jn,\omega} {\hat{b} }^\dagger_{jn} + \sum_{\tilde{j}\tilde{n}}\zeta_{\tilde{j}\tilde{n},\omega} \hat c_{\tilde{j}\tilde{n}}  + \theta^*_{\tilde{j}\tilde{n},\omega} {\hat{c} }^\dagger_{\tilde{j}\tilde{n}}, 
\eeqr 
where we have only discretized the modes in the {\it out} region which is sufficient for our purpose and expressions with and without tildes are again belong to the {\it interior} and {\it exterior} of the black hole event horizon. Using  this  and the corresponding expression for ${\hat a_{\omega}}^\dagger$  we  could   write
 $: T_{\pm\pm} (x) :_{in} $  as  a  sum of  quadratic  terms in the operators $\lbrace\hat b_{jn} , {\hat{b}}^\dagger_{jn},   \hat c_{jn},  {\hat{c} }^\dagger_{jn}\rbrace$   which    act  in  a   simple form on the  states $ |F\rangle^{int}\otimes|F\rangle^{ext} $.

Unfortunately    such  calculation  turns out to be  extremely difficult,   even when one  takes  some   simple  form for the  functions   $\lbrace  w_{nj}\rbrace $   (we   had  considered  the case   where  those  functions are just constants).

As evident, so far, with the above analysis we can write down a formal general expression of the backreacted metric in presence of wavefunction collapse. As we have mentioned, the normal ordering contributes to $\Omega$ in \eqref{soln} via \eqref{solnF} and \eqref{solnG}. Of course, if we consider \eqref{CSL-QFT3}, the CSL evolved state of the {\it in} vacuum for the right movers, we can no longer neglect the normal ordered part and an appropriate expression for them should be
\beqr
\langle \psi_{CSL} |: T_{\pm\pm}(x) :_{in}| \psi_{CSL} \rangle &=& t_{\pm\pm}(x) 
\label{tpm}
\eeqr 
and we shall end up with a backreacted spacetime, given by
\beqr
\chi &=& \Omega=-\frac{\Lambda^2x^+x^-}{\sqrt{\kappa}}-\frac{\sqrt{\kappa}}{4}\ln(-\Lambda^2x^+x^-)-\frac{m}{\Lambda\sqrt{\kappa}x_{0}^+}(x^+-x_{0}^+)\theta(x^+-x_{0}^+) \nonumber\\
&& + \frac{1}{\sqrt{\kappa}}\left( \int^{x^+} dx^{'+} \int^{x^{'+}} dx^{''+} t_{++} + \int^{x^-} dx^{'-} \int^{x^{'-}} dx^{''-} t_{--}  \right) 
\eeqr
where new undetermined functions $t_{\pm\pm}$ appear due to the CSL excitation of the vacuum state. One shortcoming of not having the explicit expression of the normal order expectation value in {\it general CSL state} \eqref{CSL-QFT3} is that we cannot compute an exact backreacted spacetime which needs to be a continuous change over the standard RST space-time. However, for completeness, in Appendix B, we explicitly compute the backreacted and modified RST space-time for a single, GRW type of collapse event. Even in this situation the metric can be glued continuously on the collapse hypersurface. The case of continuous collapse would need to generalize this situation for multiple collapse events and gluing the spacetime for each of those events over the family of such collapse hypersurfaces. This task, however, is beyond the scope of this paper.

The point is that    we know  from construction of the CSL  type of evolution,   together  with our  assumption that $\lambda$   depends on curvature \eqref{lamr}  and  diverges as  one approaches the singularity and that  as one   considers an hypersurface very  close to the singularity, the state there  would  have  collapsed to a state  with definite  occupation number, giving
 \beq
 | \psi_{CSL} \rangle = N C_{F_0}|F_0\rangle^{int}\otimes|F_0\rangle^{ext}
 \eeq
  where  $F_0$  stands  collectively for a  complete   set of  the occupation numbers  in each mode  i.e.
   $\lbrace F_0^{nj} \rbrace $.  This state    is     associated  with a   flux of positive  energy  towards future null infinity  given  by  $ E_0  \approx \sum_{nj} F_0^{nj} \omega'_{nj}$  where   $\omega'_{nj}$ is the     mean  frequency  of the mode  ${n,j}$.   The  state is  also  associated  with a  corresponding (equal)   negative  energy   flux into the black hole.  Therefore, associated  with the  black hole  at late times  we would have a total   energy  corresponding  the   mass  associated  with the  pulse that led to the formation of the black  hole
 $M-E_0$    while  the    Hawing  radiation  would have  carried to   $\cal{I}^+ $  the   energy  $E_0$.  If   quantum gravity cures the singularity  without  leading to  large  violations of  energy-momentum,   and   if we ignore the  actual violation of energy-momentum associated  with CSL {\footnote{with the idea  that   eventually a realistic  calculation will be done with   a fully relativistic collapse theory \cite{BedinghamTh, RCM} with no  violation of  energy conservation akin \cite{Bedingham}.}}  then the  {\it thunderpop}  which  we know  in this model, associated with  the  final and complete evaporation of   the black hole,  would have to carry  the extra  energy   in the amount   $ E^{thunderpop} =M-E_0$.
 
 Thus  taking into  account the   assumption that quantum gravity resolves the   singularity, and  the  above   characterization of the  {\it thunderpop}  can   describe the full evolution of the initial state   from  asymptotic  past   to the   null future infinity  as  starting  with the initial  state   
 \begin{equation}
|0\rangle_{in}^{R}\otimes|Pulse \rangle^{L} =\sum_{F}C_{F}|F\rangle^{int, R}\otimes|F\rangle^{ext,R}\otimes|Pulse \rangle^{L},
\end{equation}
  transforming  as  a  result  of the CSL   collapse into
  \begin{equation}
 |F\rangle^{int, R}\otimes|F\rangle^{ext,R}\otimes|Pulse \rangle^{L},
  \end{equation}  
 on an  hypersurface  extending to  null infinity  but  staying behind    and  very  close to the singularity ( or QG region)   and  eventually  leading to the  state  at ${\cal I}^+$  given  by
  \begin{equation}
|F\rangle^{ext}\otimes|thunderpop, M-E_0 \rangle.
  \end{equation}  
 The point however is that this state is  undetermined  because we can  not  predict  which realization of the stochastic functions $w_{nj} (t)$ will  occur in   a  specific  situation.

\section{Recovering the  thermal  Hawking  radiation}
\label{hawking}
 
 In order to  deal with  the  indeterminacy, brought in by the stochasticity of the CSL  evolution, it is  convenient to consider    an ensemble  of  systems,  all prepared in the same  initial  state  and,  described by the  pure  density matrix
\begin{equation}
\rho_{0}=|\Psi_i\rangle\langle\Psi_i|,
\end{equation}
which can, by using \eqref{instate}, be written as
\begin{equation}
\rho_{0}=\rho(\tau_{0})\otimes|Pulse^L\rangle\langle Pulse^L |,
\end{equation}
where
\begin{equation}
\rho(\tau_{0})=\sum_{F,G}C_{F, G}|F\rangle^{int}\otimes|F\rangle^{ext}\langle G|^{ext}\otimes\langle G|^{int},
\end{equation}
and $C_{F, G}$ are determined by the Bogolyubov coefficients between various mode functions. Now time evolution according to the CSL dynamics suggests \cite{Modak:2014qja} 
\begin{equation}
\rho(\tau)=\mathcal{T}e^{-\int_{\tau_{0}}^{\tau}d\tau'\frac{\lambda(\tau')}{2}\sum_{n,j}[N_{n,j}^L-N_{n,j}^R]^2}\rho(\tau_{0}).
\end{equation}
In the late time limit we have
\begin{equation}
\rho(\tau)=\sum_{F, G}e^{-\frac{\pi}{\Lambda}(E_{F}+E_{G})}e^{-\sum_{n,j}(F_{n,j}-G_{n,j})^2\int_{\tau_{0}}^{\tau}d\tau'\frac{\lambda(\tau')}{2}}|F\rangle^{int}\otimes|F\rangle^{out}\langle G|^{out}\otimes\langle G|^{int}.
\end{equation}
Taking into account the dependence of $\lambda$ on $R$ and the fact that inside the black hole and the foliation makes $R$ a function of $\tau$, we conclude that for the evolution under consideration, $\lambda$ becomes effectively that function of $\tau$. Noting the manner in which $\lambda(\tau)$ in (\ref{lamr}) depends on $R$ we conclude that the integral diverges near the singularity. Therefore the only surviving terms are the diagonal ones and thus very close to the singularity we will have,
\begin{equation}
\lim_{\tau\rightarrow\tau_{s}}\rho(\tau)=\sum_{F}e^{-\frac{2\pi}{\Lambda}E_{F}}|F\rangle^{int}\otimes|F\rangle^{out}\langle F|^{out}\otimes\langle F|^{int}.
\end{equation}
Next we explicitly include the left moving pulse, so that the complete density matrix very close to the singularity is given by
\begin{equation}\label{final}
\lim_{\tau\rightarrow\tau_{s}}\rho'(\tau)=\sum_{F}e^{-\frac{2\pi}{\Lambda}E_{F}}|F\rangle^{int}\otimes|F\rangle^{out}\langle F|^{out}\otimes\langle F|^{int}\otimes|Pulse\rangle\langle Pulse|.
\end{equation}
Note that $E_{F}$ represents the energy of state $|F\rangle^{ext}$ as measured by late time
observers. The operator given by eq. (\ref{final}) represents the ensemble when the
evolution has almost reached the singularity.

Next  by  taking into account the   effects  of the  quantum gravity region, {\footnote{That, would include for instance, the effects represented by the {\it thunderpop} in RST model.}}  on   each one  of the  components of the   ensemble   characterized  by the  above  density  matrix,  we  can  write the   corresponding density matrix  for the ensemble at  $ {\cal I}^+$  namely:
\begin{equation}\label{final2}
\rho'( {\cal I}^+) =\sum_{F} e^{-\frac{2\pi}{\Lambda}E_{F}}
|F\rangle^{out} \langle F|^{out}\otimes
|T, M-E_0(F) \rangle \langle T, M-E_0(F)|,
\end{equation}
where $T$ stands for the {\it thunderpop} and the state has energy $M-E_0(F)$.

As far as the information is concerned  we  do not believe that  the
correlations   between the {\it thunderpop}  energy   and  that  in  the  early
  parts  of the Hawking  radiation   can    be of any help   in
restituting a unitary relation between the initial and   final states.
This  is  because   the   same  considerations concerning the possibility
that a remnant  might  help  in this regard,   apply to the  {\it thunderpop}.
That is,    the amount of energy  available to the {\it thunderpop} is expected
to be rather small and to a large extent independent of the  initial mass
of the  matter  that   collapsed to form a  black hole,  and  thus  for
large enough  initial masses,  the    overwhelming  part of the initial
energy  would be emitted in the form of Hawking radiation. The  small, and
  essentially fixed,   amount of energy   available  to the {\it thunderpop}
is  not   expected to be  sufficient for the excitation of the
arbitrarily  large     number of  degrees  of freedom  necessary to
restore unitarity to en  entangle    the Hawking  radiation   and  {\it thunderpop} state.   Thus the  resulting picture   emerging from the present work is  consistent with  a  full loss  of information    during the
evaporation of a black hole  that was present in our previous treatments.

\section{Discussion}

The  black hole information problem continuous to be  a topic  attracting  wide spread interest. 
A  proposal to deal with the issue in  a  scheme    which  unified  it with the general  measurement   problem  in quantum  mechanics,   has  been advocated  in  \cite{Okon0, Okon1}  and  initially   studied in detail in \cite{Modak:2014qja, Modak:2014vya}.  
 Those initial proposals  left out    two important   aspects: relativistic covariance of the proposal  and  the issue of   back reaction. The former was  explored  in  \cite{Bedingham}  while the later   is the object of the   present  work.      

We have  studied  the incorporation of   spontaneous  collapse dynamics  into  the back reaction of    an  evaporating black hole   using the RST model. We have  shown in detail  how  a single  collapse leads to the modification of  the space-time and discussed  in general  how  the full  continuous  collapse  dynamics  might be used  in this case  and   might be  expanded to deal with  more realistic    Black hole models.\vspace{1cm}

{\bf Acknowledgments}: Authors thank Leonardo Ort\'iz for useful help at the initial stage of the work. Part of the research of SKM was carried out when he was an International Research Fellow of Japan Society for Promotion of Science. He is currently supported by a start up research grant PRODEP, from SEP, Mexico. DS is supported by CONACYT project 101712, Mexico, and by PAPIIT--UNAM grant IG100316, Mexico.

\section{Appendix A: The renormalized  Energy-momentum tensor}

Here we  review, and to  a certain degree  clarify, the  method to calculate $\bra{\Psi}| T_{ab} |{\Psi} \ket$ using the properties of CGHS model and a result for the renormalized energy-momentum tensor as found by Wald \cite{wald-ren}. 

Let us start  with  the 
expression for the renormalized energy-momentum tensor  obtained in \cite{wald-ren}. The result   is    described using Penrose's  abstract  index  notation for clarity.  It applies to  a   two  dimensional spactime  where the metric  is conformally related to the flat metric so that $g_{ab} = \Omega^2 \eta_{ab }$  and   where,  in the past  the  metric (is  or  approaches  asymptotically)  the flat  Minkowski metric,  so that  there $\Omega^2 = 1$.
It  offers  an  expression of the   renormalized    expectation value of the energy-momentum tensor,  in   terms of the  derivative operator  $\nabla^{(\eta)}_a$ associated  with   the flat  metric   $\eta_{ab}$.
 The   expression  is:
\beqr
\langle \Psi | T_{ab} | \Psi \rangle &=& \langle \Psi |:T_{ab}:_{in}| \Psi \rangle + \frac{1}{12\pi} (\nabla^{(\eta)}_{a}\nabla^{(\eta)}_{b} C - \nonumber\\
&&  \eta_{ab} \eta^{cd} \nabla^{(\eta)}_{c} \nabla^{(\eta)}_{d} C -\nabla^{(\eta)}_a C \nabla^{(\eta)}_b C + \frac{\eta_{ab}}{2} \eta^{cd}\nabla^{(\eta)}_{c} C \nabla^{(\eta)}_{d} C), \label{rem}
\eeqr
where the first expression on the right hand side is the normal ordering with respect to the  construction of the  quantum  field  theory  that leads to the 
{\it in} vacuum, $\eta_{ab}$ is the flat metric,
and $C = \ln\Omega$. In  \cite{wald-ren},  the  derivation of  this expression   is obtained using an axiomatic approach to renormalize the stress tensor.  We emphasize that the  above  expression is  fully covariant when  all the objects are   properly understood,
(in particular $ \nabla^{(\eta)}_a $    is derivative operator  associated  with the  flat metric  $\eta_{ab}$),  and it is valid  in all regions of  space-time and not only in the flat region   in the past (where the  expression would simply be $  \langle \Psi |:T_{ab}:_{in}| \Psi \rangle $ as   there  $C =$ constant).

Note that  as \eqref{rem} is expressed in terms of the  derivative operator  associated  with the  flat metric  $\eta_{ab}$,   thus it can be   simply  written in    terms of the  ordinary    derivative  operator     $\partial^{(y)} $ associated    with Minkowski   coordinates  $y^\mu$  in which one  can  write the  flat    metric    components  as $ \eta_{\mu \nu}$ ( i.e.   $  \eta_{ab} =\eta_{\mu \nu} dy^\mu_a {dy^\nu}_b   = -dy^0_a dy^0_b + dy^1_a {dy^1}_b $  (because this operator  coincides  with  the covariant  derivative operator associated with the flat  metric). That is,   we   can  write  the  expression  in terms of the explicit   components  in these  coordinates as:

\beqr
\langle \Psi | T_{\mu\nu} | \Psi \rangle &=& \langle \Psi |:T_{\mu\nu}:_{in}| \Psi \rangle + \frac{1}{12\pi} (\partial^{(y)}_{\mu}\partial^{(y)}_{\nu} C - \nonumber\\
&&  \eta_{\mu\nu} \eta^{\alpha\beta} \partial^{(y)}_{\alpha} \partial^{(y)}_{\beta} C -\partial^{(y)}_\mu C \partial^{(y)}_\nu C + \frac{\eta_{\mu\nu}}{2} \eta^{\alpha\beta}\partial^{(y)}_{\alpha} C \partial^{(y)}_{\beta} C), \label{rem2}
\eeqr
   where the   derivative operators  are  just partial  derivatives  with respect to the coordinates  $y^{\mu}$  above.
 In order to use this expression at an arbitrary point, where the space-time is  expressed  in    other   generic  coordinates one needs to rely  on  in appropriate   covariant  form  \eqref{rem}.

 When   evaluating the   expectation value  in  any state  we  must   ensure  that  we  use   the normal ordering with respect to the {\it in}  quantization, so that the first term on the r.h.s will be zero if we chose $|\Psi\rangle$ to be the {\it in} vacuum. 
 
 In our calculations,  we will be using  the relationship between the  derivative operators   corresponding to  the  flat  metric   and  that corresponding to  the   general metric.
Recall  that  relationship   between two derivative operators   is  represented  by is a tensor of type (1,2)  denoted  $C_{ab}^c$  which   specifies    how  it  acts  on a dual vector field $A_b$ \cite{wald-book}
\beqr
\nabla_{a} A_{b} = \nabla'_{a}A_{b} - C_{ab}^{c} A_{c}. \label{reldo}
\eeqr
Such   expression  is   of course  valid in  all   coordinate systems.
 In order to use \eqref{rem} for computing \eqref{rem2} we have to write  the ordinary derivative operator associated with the asymptotic past ``{\it in}" coordinates  which as  we  noted  is the same  as   the  covariant  derivative   operator  $\nabla^{(\eta)}_a$ associated  with the flat  metric  $\eta_{ab}$  in  terms of the  derivative operators  associated  with the  coordinates  that  cover the  whole  space-time (that is  the coordinates $ x^{\pm}$). We  denote these latter derivative  operators as $\partial_a^{(x)}$.

Next we    compute the  $C_{ab}^{c}$   which becomes the Christoffel symbol  relating the   covariant  derivative   operator  $\nabla^{(\eta)}_a$   with    the  ordinary    derivative operator $\partial_a^{(x)}$.  In the  $x$ coordinate basis   it is
\beq
\Gamma_{\mu\nu}^{\rho} =  \frac{1}{2}g^{\rho\sigma} (\partial_{\mu}g_{\sigma\nu} + \partial_{\nu}g_{\mu\sigma} -\partial_{\sigma}g_{\mu\nu}) \label{cd}.
\eeq
Note that here the Christoffel symbol is a tensor field associated with the derivative operator $\nabla_a^{(\eta)}$ and  the coordinate chart   $x^{\mu}$  associated  with the ordinary derivative operators $\partial^{(x)}_a$.

The  {\it in}   flat  metric  can be  expressed    in the  global coordinates $x^+,  x^- $    as  $ds^2_{\eta_{\mu\nu}} = \frac{dx^+ dx^-}{ (-\Lambda^2 x^+ x^-)} $   and   can   also  be  expressed in the {\it in} coordinates  as $ \eta_{\mu \nu} dy^\mu {dy^\nu} $.
The     relation  between  the  coordinates   $x^{\pm}$  and  the   coordinates   $ y^{\pm} $    is   given by  $dy ^{\pm}=d x^{\pm}/  \Lambda x^{\pm} $ while  $  dy^+ =  dy^0 + dy^1$    and  $dy^- =   dy^0 - dy^1$. Thus    we have  the metric  components   $ g_{x^+ x^+} =g_{x^- x^-} =0 $  and  $ g_{x^+ x^-} =  g_{x^- x^+} =  -(2 \Lambda^2  x^+ x^-)^{-1} $.

Next  we   find  the  appropriate  expression for the   conformal factor  relating  the flat and curved metrics. For that let us recall the spacetime metric in the conformal gauge is
\beqr
ds^2 &=& -e^{2\rho} dx^+ dx^-, \nonumber\\
&=& e^{2\rho} (-\Lambda^2 x^+ x^-) ds^2_{\eta_{\mu\nu}}, \label{fmt}
\eeqr
where is the flat linear-dilaton spacetime and the conformal factor or subsequently $C$ is found to be
\beqr
\Omega^2 &=& e^{2\rho} (-\Lambda^2 x^+ x^-), \nonumber\\
\implies C &=& \ln\Omega = \rho + \frac{1}{2}\ln(-\Lambda^2x^+ x^-). \label{cf}
\eeqr
Using \eqref{reldo}, \eqref{cd}, \eqref{fmt} and \eqref{cf}, a simple calculation yields
\beqr
\bra{\Psi}| T_{x^\pm x^\pm} |{\Psi} \ket &=& -\frac{\hbar}{12\pi} \left((\p_{x^\pm}\rho)^2 - \p_{x^\pm}^2\rho - \frac{1}{4{x^\pm}^2}\right) + \bra{\Psi}|: T_{x^\pm x^\pm}:_{in} |{\Psi} \ket, \label{rempp}\\
\bra{\Psi}| T_{x^+ x^-} |{\Psi} \ket &=& -\frac{\hbar}{12\pi}\p_{x^+}\p_{x^-}\rho + \bra{\Psi}|: T_{x^+x^-}:_{in}|{\Psi} \ket \label{rempm}.
\eeqr

 As  it  is  well known \cite{davies}, \cite{wald78}, the  covariant behavior of the renormalized stress tensor (which is a direct consequence of semi-classical Einstein equations) requires  a   specific nonzero trace of  the expectation value of the  energy-momentum tensor that should  be  in fact   state independent.  That  is :
\beqr
g^{\mu\nu} \langle \Psi | T_{\mu\nu} | \Psi \rangle = \frac{\hbar R}{24\pi}. \label{tran}
\eeqr
In the conformal gauge where the only non-vanishing metric components are $g^{x^+ x^-}$ one can easily find the off-diagonal components of renormalized energy-momentum tensor $\langle \Psi | T_{x^{\pm}{x^\mp}} | \Psi \rangle$ which matches with \eqref{rempm} given the  fact that  $\bra{\Psi}|: T_{x^+x^-}:_{in}|{\Psi} \ket $
 must vanish  (this  vanishing is  a consequence of  the conservation law $\nabla^\mu \bra{\Psi}|T_{\mu\nu }|{\Psi} \ket  =0 $). This is a direct consequence of the fact that trace of the energy-momentum  tensor which appears in \eqref{tran} is independent of the state of the quantum field.

Thus we now have the  explicit expressions for various components of the renormalized energy-momentum tensor that appear in semi-classical Einstein equations \eqref{gpmmb} and \eqref{gppmb}, given by
\beqr
\bra{\Psi}| T_{x^\pm x^\pm} |{\Psi} \ket &=& \frac{\hbar}{12\pi} \left(\p_{x^\pm}^2\rho -(\p_{x^\pm}\rho)^2 - \frac{1}{4{x^\pm}^2}\right) + \bra{\Psi}|: T_{x^\pm x^\pm}:_{in} |{\Psi} \ket, \label{remppf}\\
\bra{\Psi}| T_{x^+ x^-} |{\Psi} \ket &=& -\frac{\hbar}{12\pi}\p_{x^+}\p_{x^-}\rho  \label{rempmf}.
\eeqr

As a  check  on the above expressions  we consider  them  for  the {\it in} vacuum state in the linear dilaton vacuum region,  where components of the renormalized energy-momentum tensor should vanish. The conformal factor in linear dilaton vacuum $e^{2\rho(x^\pm)} = -\frac{1}{\Lambda^2 x^+ x^-}$ which implies that the first term in \eqref{remppf} vanishes and the normal ordered part is trivially zero since we have chosen $|\Psi\rangle$ to be the vacuum state. Similarly it is easy to see that \eqref{rempmf} vanishes as well. 
This provides one  consistency check for  the expressions of  the renormalized energy-momentum tensor.

\section{Appendix B: The backreacted spacetime with GRW type collapse}

Here we want to explicitly calculate the backreacted spacetime assuming a single collapse event of GRW type \cite{GRW:86} on one of the collapse hypersurfaces chosen stochastically. Also, we would only consider a situation where the right-moving modes are  subjected to collapse. In this situation the  backreacted metric due to collapse has the following form
\beqr
\chi &=& \Omega=-\frac{\Lambda^2x^+x^-}{\sqrt{\kappa}}-\frac{\sqrt{\kappa}}{4}\ln(-\Lambda^2x^+x^-)-\frac{m}{\Lambda\sqrt{\kappa}x_{0}^+}(x^+-x_{0}^+)\theta(x^+-x_{0}^+) \nonumber\\
&& + \frac{1}{\sqrt{\kappa}} \int_0^{x^+} dx^{'+} \int_0^{x^{'+}} dx^{''+} t_{++} 
\label{dmetr1}
\eeqr
where $t_{++}$ is a state dependent function, defined in \eqref{tpm}, and it vanishes if and only if we chose the quantum state to be the ``in'' vacuum. In that case we have the RST spacetime. However, if the state is different than vacuum, such as when  the wavefunction  is  modified  by  a  the  collapse dynamics, we must find out the  new spacetime.

We  will   consider  here a single collapse event,   associated with a  certain  space-like  hypersurface,  which  we     take  here  as given    by   one of the  hypersurfaces of the folliation  introduced in section    V.D.  That   is  the hypersurface that  corresponds to say to a specific   value   of $R$   say   $ R= R_c$ (which  in the distant exterior  region  is  matched to something else  as    shown in   Fig. 3).    In order to further simplify matters   we  will  chose  chose    $t_{++}$  to be    proportional to a (localized) function of compact support along $x^+$. 
This   is   further   motivated  by the  form of the  collapse operators   which are  associated  to the  modes  $ n, j $   which as  we   know  are highly localized.     However   as we noted,   the meaningful use of   the semi-classical setting, requires   that   the precise form of the collapse operators  should  be   such as to ensure the   Hadamard  nature of the states  that result from the collapse dynamics.  

  We represent the   fact   that $t_{++}$ is nonzero only to the future of the collapse hypersurface characterized by the equation $x^+ = f_{R_c}(x^{-})$,  by   including  a theta function. Keeping these   considerations in mind we have
\beqr
t_{++} (x^+, x^-) = \epsilon h(x^+) \Theta [x^+ - f_{R_c}(x^-)] ,\nonumber \\ \label{tp}
\eeqr
where $\epsilon$ is small number, $x^+ - f_{R_c}(x^-) =0$ specifies the collapse hypersurface and the theta function makes sure that the energy-momentum tensor vanishes to all the points past to the collapse hypersurface\footnote{Note that the foliation is made here with respect to $R$ which is monotonically increasing with respect to the Cauchy slicing.}. The resulting backreacted spacetime is then found by putting \eqref{tp} in  \eqref{dmetr1} and integrating.

As an example,   where  the calculation can be made  explicitly we considered 
the function
\begin{multline}
h(x^+) =
\left
\{
\begin{array}{ll}
 0  & \hspace{20pt} x^+\le c \\
        \alpha (x^+ - c) &  \hspace{20pt} c\le x^+ \le c+b/2\\
        \alpha b -\alpha (x^+ - c)  &\hspace{20pt}  c+b/2\le x^+ \le c+b\\
        0 & \hspace{20pt} b+c<x^+
\end{array}
\right.
\label{hx}
\end{multline}
and checked that the resulting  space-time    metric  is  smooth,    everywhere    except  on the  collapse  hypersurface where it is only   continuous.  The spacetime is  modified  only to the  future of the  support of $t_{++}$.

\begin{figure}[h]
\centering
\includegraphics[scale=0.38]{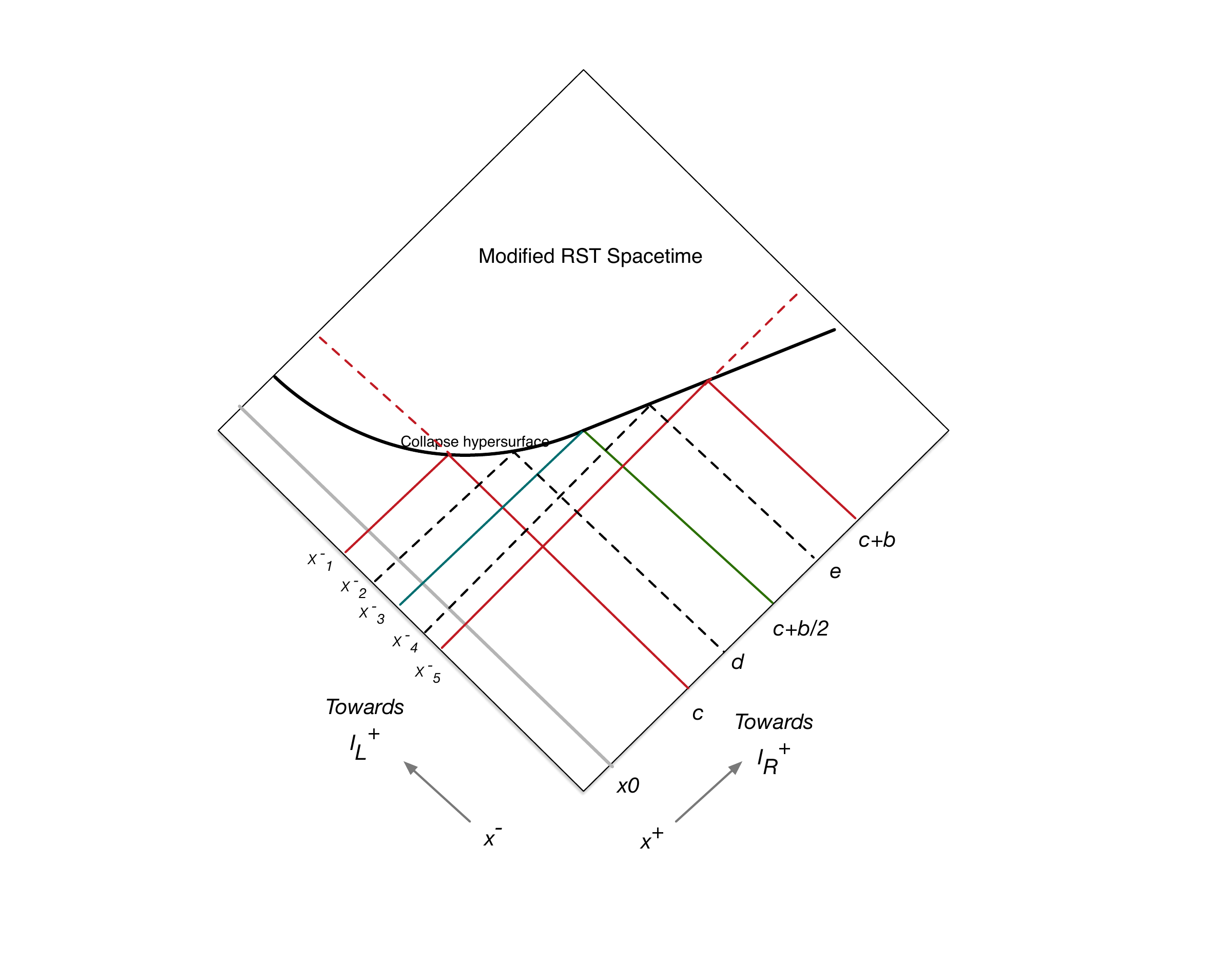}
\caption{Modified RST spacetime due to the collapse of wavefunction in Kruskal frame. The energy-momentum tensor due to the collapse of the ``in'' vacuum state has support in the region $c\le x^+ \le c+b$ which modifies the spacetime after the collapse hypersurface. The modification to RST spacetime is the intersection of the future light cone of the point $(x^+=c, x^-=x^-_1)$ and the causal future of all the points on the collapse hypersurface.}
\label{mrst}
\end{figure}

The  analysis proves that as we move along the $x^-$ axis  ( with fixed $x^+$)  the metric changes nontrivially as  one  crosses  the hypersurface, but, the change is always continuous. 

As  one moves  along  a  line of  fixed $x^-$  by changing $x^+$,   the modification appears  only to the future of the collapse hypersurface and  change is continuous   The pictorial description of these results is depicted  in Fig. 4.

 By observing    Fig. 4  and Fig. 3  we    note that    the collapse  can  result in modifications  of the  metric  and  quantum state  at   asymptotic infinity,  in the  black hole region and   on the   {\it thunderpop}.
 
 The  actual changes  will of course  depend  on the  specific  realization of the stochastic  parameters/functions  that control the collapse  evolution,  as these will determine the    actual  state of the quantum field  that  results  from the collapse.  The treatment presented here   is  limited to a single  instantaneous  collapse  event  and the  treatment involving  the continuous  and multi-mode  CSL  dynamics,  presented in     section     V. B.,  is  more complicated  but   can be  generalized  as  a continuous  version of what  we  showed here. 
   

\end{document}